\documentclass[aps,pre,twocolumn]{revtex4-2}
\usepackage{amsfonts,amssymb,amsmath,bm}
\usepackage{graphicx}% Include figure files
\usepackage{hyperref}% add hypertext capabilities
\usepackage{color}
\usepackage{physics}
\usepackage[dvipsnames]{xcolor}

\begin{document}

\title{Conformational statistics of non-equilibrium polymer loops\\in Rouse model with active loop extrusion}
\date\today

\author{Dmitry Starkov}
\author{Vladimir Parfenyev}
\author{Sergey Belan}
\email{sergb27@yandex.ru}
\affiliation{Landau Institute for Theoretical Physics, Russian Academy of Sciences, 1-A Akademika Semenova av., 142432 Chernogolovka, Russia}
\affiliation{National Research University Higher School of Economics, Faculty of Physics, Myasnitskaya 20, 101000 Moscow, Russia}

\begin{abstract}
Motivated by the recent experimental observations of the DNA loop extrusion by protein motors, in this paper we investigate the statistical properties of the growing polymer loops within the ideal chain model. The loop conformation is characterised statistically by the mean gyration radius and the pairwise contact probabilities.  It turns out that a single  dimensionless parameter, which is given by the ratio of the loop relaxation time over the time elapsed since the start of extrusion,  controls the crossover between near-equilibrium and highly non-equilibrium asymptotics in statistics of the extruded loop.
Besides, we show that two-sided and one-sided loop extruding motors produce the loops with almost identical properties.
Our predictions are based on two rigorous semi-analytical methods accompanied by asymptotic analysis of slow and fast extrusion limits.
\end{abstract}

\maketitle

\section{Introduction}

According to the loop extrusion model, the nanometer-size molecular machines organize chromosomes in nucleus of living cells by producing dynamically expanding chromatin loops \cite{Alipour_2012,Goloborodko_2016}. The molecular dynamics simulations of chromatin fiber subject to loop extrusion allow to reproduce the in vivo 3D chromosome structures and explain the origin of interphase domains observed in experimental Hi-C data \cite{Sanborn_2015,Nuebler_2018,Gibcus_2018,Banigan_2020b}. Importantly, being originally proposed as a hypothetical molecular mechanism, the loop extrusion process has been  observed in the recent single-molecule experiments in vitro \cite{Ganji_2018,Golfier_2020,Kong_2020}. Namely, these experimental studies showed that the Structural Maintenance of Chromosome (SMC) protein complexes, such as cohesin and condensin, can bind to chromatin and extrude a loop due to the ATP-consuming motor activity.

From the statistical physics point of view, chromatin fiber subject to loop extrusion is an intriguing example of non-equilibrium polymer system. While we have a (comparably) satisfactory theoretical picture of equilibrium macromolecules \cite{DeGennes_1979,Doi_1988, GKh_1994}, the statistical physics of non-equilibrium polymers is a territory of many open questions \cite{Grosberg_2016,Huang_2018,Vandebroek_2015,Samanta_2016,Sakaue_2017,Osmanovic_2017, Winkler_2017, Lowen_2018, Osmanovic_2018, Chaki_2019, Put_2019, Anand_2020}. A large research interest around this field is motivated by ongoing advances in development of experimental techniques providing unprecedented insights into structure and dynamics of biological polymers in living cells \cite{Dekker_2002, Bolzer_2005, Lieberman_Aiden_2009, Joyce_2012, Nagano_2013, Shachar_2015, Fraser_2015, Kind_2015, Berkum_2010, Oudelaar_2018, Tavares_Cadete_2020, Oomen_2020,Krietenstein_2020,Oudelaar_2018}.

In attacking the problem of chromatin modeling in the view of newly established (but conceptually old \cite{Riggs_1990}) loop extrusion mechanism it is natural to start with the following simple question: how does the incorporation of active loop extrusion change the  properties of the canonical polymer models? Here we take the first step on this research program. Adopting the Rouse model of an ideal polymer chain (see, e.g., \cite{Doi_1988, GKh_1994}), we explore how the conformational properties of the dynamically growing polymer loops differ from that of the static equilibrium loops.
Our analysis allows to predict the effective size of the extruded loop, measured in terms of the gyration radius, and contact frequency between monomers inside the loop in their dependence on the extrusion velocity.

%In this paper we develop an  analytical model of loop extrusion with the aim of describing the conformational statistics of  non-equilibrium  loops.
%Namely we consider a Rouse polymer - a  chain of beads connected by harmonic springs - subject to the loop extrusion process.

\section{Model formulation}
\label{sec: model}

\begin{figure}[t]
\centering{\includegraphics[width=\linewidth]{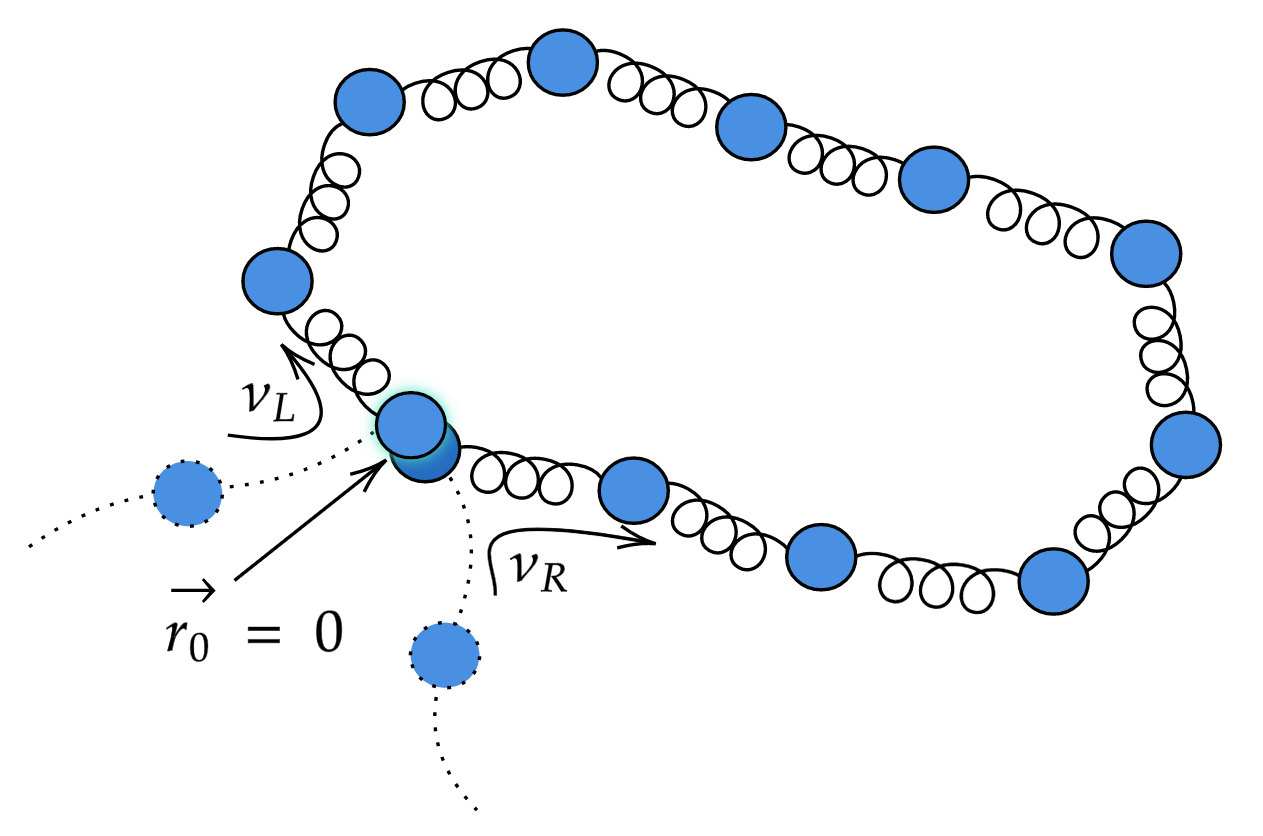}}
\caption{Growth of polymer loop via two sided extrusion process. The extrusion velocities $\nu_L=\frac{1-q}{2}\nu_0$ and $\nu_R=\frac{1+q}{2}\nu_0$ represent the rates at which the LEF adds new beads at the left and right arm of the loop, respectively, and $q$ is the so-called symmetry score introduced in Ref \cite{Golfier_2020}. The choice $q=1$ corresponds to the pure one-sided extrusion, while at $q=0$ we deal with perfectly symmetric two-sided loop growth.}\label{Fig: 1}
\end{figure}

Consider a long chain of beads connected by the identical harmonic springs and placed into a thermal bath.
We assume that a single loop extruding factor (LEF) loads a polymer chain at the time moment $t=0$ and starts producing a progressively growing loop. In general, extrusion may occur at left and right sides at different rates $\nu_L$ and $\nu_R$ (see Fig.~\ref{Fig: 1}), but for now we consider the case of pure one-sided extrusion that corresponds to the unit symmetry score $q=1$ (i.e. $\nu_L=0$ and $\nu_R=\nu_0$) and return to discussion of the two-sided extrusion in the last section.
Then,
the number of beads in the loop as a function of time $t$ elapsed since the start of extrusion process is given by $N(t)=1+[\nu_0t]$, where  $\nu_0$ is the rate at which the LEF operates beads and $[\dots]$ denotes the integer part of the number. It is convenient to label the beads in the loop by integer numbers $0,1,\dots, N(t)-1,N(t)$, where index $0$ corresponds to the loading site of the LEF.

The stochastic dynamics of the chain is governed by interplay of the inter-beads attraction forces, thermal noise, and the loop extrusion activity. To make this problem analytically tractable, in what follows we will assume that the LEF is fixed in the origin of the Cartesian system of coordinates. One, thus, obtains a loop that is pinned at one point and grows via addition of the new beads at $\vec r=0$ with the constant rate of $\nu_0$. The dynamics of the system during a time interval between addition of new beads is described by the following set of linear equations
\begin{equation}
\begin{aligned}\label{"eq:1"} %\label{"eq:6"}
&{\vec r}_{0}=0,\\
&\dot{\vec r}_{1} = \frac{k}{\zeta}(\vec r_{2}+\vec r_{0}-2\vec r_{1}) +\frac1\zeta\vec \xi_1(t),\\
&\dot{\vec r}_{2} = \frac{k}{\zeta}(\vec r_{3}+\vec r_{1}-2\vec r_{2}) +\frac1\zeta\vec \xi_2(t),\\
&\dots\\
&\dot{\vec r}_{N(t)-1} = \frac{k}{\zeta}(\vec r_{N(t)}+\vec r_{N(t)-2}-2\vec r_{N(t)-1}) +\frac1\zeta\vec \xi_n(t),\\
&{\vec r}_{N(t)}=0,
\end{aligned}
\end{equation}
%\begin{eqnarray}
%&&{\vec r}_{0}=0, \label{"eq:1"} \\
%&&\dot{\vec r}_{1} = \frac{k}{\zeta}(\vec r_{2}+\vec r_{0}-2\vec r_{1}) +\frac1\zeta\vec \xi_1(t),\\
%&&\dot{\vec r}_{2} = \frac{k}{\zeta}(\vec r_{3}+\vec r_{1}-2\vec r_{2}) +\frac1\zeta\vec \xi_2(t),\\
%&&\dots\\
%&&\dot{\vec r}_{N(t)-1} = \frac{k}{\zeta}(\vec r_{N(t)}+\vec r_{N(t)-2}-2\vec r_{N(t)-1}) +\frac1\zeta\vec \xi_n(t),\\
%&&{\vec r}_{N(t)}=0, \label{"eq:6"}
%\end{eqnarray}
where $\vec r_n(t)$ is the position of the $n$-th bead, $\xi_n(t)$ is the Langevin force, $k$ is the spring elasticity, $\zeta$ is the friction coefficient of a bead, and the dot denotes the time derivative. The random forces are characterised by zero mean value $\langle \xi_{n,\alpha}(t)\rangle=0$ and the pair correlator
\begin{equation}\label{eq:noise_vp}
\langle \xi_{n,\alpha}(t_1)\xi_{m,\beta}(t_2)\rangle = 2\zeta k_BT\delta_{nm}\delta_{\alpha\beta}\delta(t_2-t_1),
\end{equation}
where $k_B$ is the Boltzmann constant, $T$ is the environment temperature, $\delta_{nm}$ and $\delta_{\alpha\beta}$ are the Kronecker delta, the Latin indices denote bead numbers, the Greek indices run over $\{x,y,z\}$, and $\delta(t)$ is the Dirac delta function.

In other words, $0$-th and $N(t)$-th beads are fixed at $\vec r=0$, while other beads move being subject to harmonic interaction forces and random noises. After $\Delta t = 1/\nu_0$ has passed, we add a new bead at the loop base, which increases the total bead number $N(t)$ until another addition. The procedure of attaching new beads is repeated over and over again.

We would like to characterize the growing loop statistically in terms of two primary metrics. First of all, it is interesting to understand how the (time-dependent) contour length $N(t)$ of the loops translates into its physical size. A measure of the latter is the radius of gyration defined as
\begin{equation}
\begin{aligned}\label{gyration_radii_def}
R_g^2(t):=\frac{1}{2N(t)^2}\sum_{n,m=0}^{N(t)-1}\langle(\vec r_{n}(t)-\vec r_m(t))^2\rangle=\\
\frac{1}{N(t)}\left(\sum_{n=0}^{N(t)-1}F_{n,n}(t)-\frac{1}{N(t)}\sum_{n,m=0}^{N(t)-1}F_{n,m}(t)\right),
\end{aligned}
\end{equation}
%\begin{eqnarray}
%\label{gyration_radii_def}
%&&R_g^2:=\frac{1}{2N(t)^2}\sum_{n,m=0}^{N(t)-1}\langle(\vec r_{n}(t)-\vec r_m(t))^2\rangle=\\
%&&=\frac{1}{N(t)}\left(\sum_{n=0}^{N(t)-1}F_{n,n}(t)-\frac{1}{N(t)}\sum_{n,m=0}^{N(t)-1}F_{n,m}(t)\right),
%\end{eqnarray}
where
\begin{equation}
\label{pair_correlator}
F_{n,m}(t)= \langle \vec r_n(t)\cdot\vec r_m(t)\rangle
\end{equation}
is the pair correlation function of the beads coordinates and angular brackets denote averaging over the statistics of thermal fluctuations.

Another interesting metric characterising the spatial conformation of the loop is the pairwise contact probability between $n$-th and $m$-th beads, which is given by
%\begin{eqnarray}
 %&&P_c(n,m;t)=\text{Prob}\left[|\vec r_n(t)-\vec r_m(t)|<a_0\right]\approx\\
 %&&\approx\frac{4}{3}\pi a_0^{3}\int p(\vec r_n,\vec r_m;t)\delta(\vec r_m-\vec r_n)d^3r_nd^3r_m=\\
 %&&%=\frac{4}{3}\pi a_0^{3}(\frac{3}{2\pi\langle R_{n,m}^2(t)\rangle})^{3/2}
 %=\sqrt{\frac{6}{\pi}}\frac{a_0^3}{\left(F_{n,n}(t)+F_{m,m}(t)-2F_{n,m}(t)\right)^{3/2}},
 %\label{eq: contact}
%\end{eqnarray}
%where $a_0$ is a cutoff contact-radius, and $p(\vec r_n,\vec r_m;t)$ is the joint probability distribution of the beads coordinates.
%In derivation of Eq.~(\ref{eq: contact}) we assumed that $a_0\ll \langle |\vec r_n(t)-\vec r_m(t)|\rangle$ and exploited the Gaussian nature of our statistical model (see Appendix ? and ? for details).
\begin{equation}
\begin{aligned}\label{eq: contact}
P_c(n,m;t)=\text{Prob}[R_{n,m}(t)<a_0]\approx\\
\frac{4}{3}\pi a_0^{3}\int P(\vec R_{n,m};t)\delta(\vec R_{n,m})d^3R_{n,m}=\\
\sqrt{\frac{6}{\pi}} a_0^3 \left(F_{n,n}(t)+F_{m,m}(t)-2F_{n,m}(t)\right)^{-3/2},
\end{aligned}
\end{equation}
%\begin{eqnarray}
%&&P_c(n,m;t)=\text{Prob}[R_{n,m}(t)<a_0]\approx\\
%&&\approx\frac{4}{3}\pi a_0^{3}\int P(\vec R_{n,m};t)\delta(\vec R_{n,m})d^3R_{n,m}=\\
%&&%=\frac{4}{3}\pi a_0^{3}P(0;t)
%=\sqrt{\frac{6}{\pi}}\frac{a_0^3}{\left(F_{n,n}(t)+F_{m,m}(t)-2F_{n,m}(t)\right)^{3/2}}, \label{eq: contact}
%\end{eqnarray}
where $a_0$ is a cutoff contact-radius, and $P(\vec R_{n,m};t)= (\frac{3}{2\pi\langle R_{n,m}^2(t)\rangle})^{3/2}\exp(-\frac{3R_{n,m}^2}{2\langle R_{n,m}^2(t)\rangle})$ is the probability distribution of the inter-beads separation vector $\vec R_{n,m}(t)=\vec r_n(t)-\vec r_m(t)$. In derivation of Eq.~(\ref{eq: contact}) we assumed that $a_0\ll \sqrt{\langle R_{n,m}^2(t)\rangle}$ and exploited %the Gaussian nature of our statistical model.
the normal form of the distribution $P(\vec R_{n,m};t)$ which is due to the linearity of our model and the Gaussian properties of the noise.

From Eqs.~(\ref{gyration_radii_def}) and (\ref{eq: contact}) we see that both radius of gyration and contact probability are expressed via the pair correlator defined in Eq.~(\ref{pair_correlator}). In Sections~\ref{sec: discrete} and~\ref{sec: continuum} we present two semi-analytical approaches allowing us to compute $F_{n,m}(t)$.

%\begin{eqnarray}
%&&P_c(n,m;t)\propto\int P(\vec r_n,\vec r_m;t)\delta(\vec r_m-\vec r_n)d^3r_nd^3r_m,
%\end{eqnarray}

%We developed two semi-analytical approaches allowing us to calculate $F(n.m;t)$, which are described in the next two sections.

%Importantly, the model with a time-independent $N$ is known to have normal statistics [Ref], and the process of particle addition could be accounted for by an ensemble of Heaviside step functions in dynamical equations, which wouldn't affect its linear properties. Thus, the overall statistics is expected to be zero-mean Gaussian as well. A rigorous derivation of this fact can be found in the Appendix.

One may expect that the loops generated by sufficiently slow extruders are reminiscent to the equilibrium  Rouse coils  whose properties are well understood (see Appendix \ref{sec: app_equilibrium }). To measure the role of the non-equilibrium nature of loop extrusion we introduce the dimensionless parameter $\sigma={\tau_{\text{relax}}}/{\tau_{\text{ext}}}$, where $\tau_{\text{relax}}=N^2 /\pi^2\gamma$ represents the relaxation time of the loop having size $N$ and characterized by the kinetic coefficient  $\gamma=k/\zeta$, and $\tau_{\text{ext}}=N/\nu_0$ is the time required to extrude this loop. Therefore,
\begin{equation}
\sigma=\frac{N\nu_0}{\pi^2\gamma} \label{eq: sigma}
\end{equation}
and since the LEF progressively enlarges the loop, %so that at long enough times $N(t) \approx \nu_0 t$,
the degree of non-equilibrium  grows with time as $\sigma(t)=\nu_0(1+[\nu_0t])/(\pi^2\gamma)$. %, $\sigma \approx \nu_0(1+[\nu_0t])/(\pi^2\gamma)$.
In Section~\ref{sec: results} we will see that typical conformation of loops characterized by sufficiently small value of $\sigma $ is nearly equilibrium, whereas loops having large $\sigma$ exhibit completely different behaviour.

%\begin{equation}
%\sigma=\frac{\nu_0^2t}{\pi^2\gamma}. \label{eq: sigma1}
%\end{equation}

\section{Discrete model: Fokker-Planck equation} \label{sec: discrete}
To start tackling the problem of obtaining $F_{n,m}(t)$ we first consider a time interval $t\in [(J-1)\Delta t, J\Delta t)$ when there are $J$ beads in the system. We also make use of the fact that the problem is isotropic, which allows us to consider only the one-dimensional case. We then rewrite dynamical equations  (\ref{"eq:1"}) in the matrix form
\begin{equation}
    \dot{\vec{x}} = \hat{A}_J \vec{x} + \frac{1}{\zeta}\vec{\xi}(t)
\end{equation}
where $\vec{x}$
%= (x_1, x_2 \ ... \ x_j)^T$
is the vector of coordinates of beads along an arbitrary Cartesian axis, and $\hat{A}_J$ is a tridiagonal Toeplitz matrix, with a lower index corresponding to the current size of the system. The zeroth bead can be safely omitted because its coordinate is fixed at the origin, so the size of this matrix is actually $J-1$. It is diagonalizable by a unitary transformation $\vec{x} = \hat{P}_J \vec{y}$ (essentially a discrete Fourier transform). Here, $\vec{y}$ is a vector of projections along so-called Rouse modes \cite{GKh_1994}.

 To avoid treating the issue of time-dependent dimensionality of $\vec{x}$ formally, we can think that $\vec{x} \in \mathbb{R}^{M}$ where $M>J$. Consequently, if there are currently $J$ beads, including the omitted one, $\hat{A}_{J}$ should be treated as a block-diagonal matrix, with a '$(J - 1) \times (J - 1)$' block acting on the non-trivial subspace of currently 'active' beads, which have already been added to the loop, and another block being an arbitrarily large identity.  The same applies to every other matrix with a lower index of $J$.

The Rouse modes evolve independently from each other
and
the marginal probability distribution ${\rho}_j (y_j, t)$ of the mode amplitude obeys the  Fokker-Planck equation \cite{Risken_1996}
\begin{equation}
    \partial_t {\rho}_j (y_j, t) = -\lambda_j \partial_{y_j} {\rho}_j(y_j,t) + D\partial^2_{y_j} {\rho}_j(y_j,t), \label{"eq:FP0"}
\end{equation}
where  $\lambda_j$ denotes the $j$-th eigenvalue of $\hat{A}_J$,  and  $D=k_B T/\zeta$ is the  diffusion constant of a single bead.
Then the joint probability density    ${\rho}_{J} (\vec{y}, t) = \prod_{j = 1}^{J} {\rho}_j(y_j, t)$
%\begin{equation}
 %   {\rho}_{J} (\vec{y}, t) = \prod_{j = 1}^{J} {\rho}_j(y_j, t), \label{eq:product}
%\end{equation}
 can be expressed as
\begin{equation}
\begin{aligned}\label{eq:propagate0}
    \rho_{J}(\vec{y}, t) = \int d\vec{y}_0 \ \rho_{J}(\vec{y}_0, (J-1)\Delta t) \\ \times  {\cal G}_{J}(\vec{y}, t-(J-1)\Delta t|\  \vec{y}_0).
\end{aligned}
\end{equation}
Here ${\rho}_j(y_j, (J-1)\Delta t)$ is the initial condition  at the moment just after the appearance of the $J$-th bead in the loop  base, and $ {\cal G}_{J}(\vec{y}, t|\  \vec{y}_0)=\prod_{j = 1}^{J} {\cal Q}_j(y_j, t| y_{j0})$ where ${\cal Q}_j(y_j, t| y_{j0})$ represents the solution of Eq.~(\ref{"eq:FP0"})  with the initial condition ${\cal Q}_j(y_j, 0| y_{j0})=\delta(y_j-y_{j0})$.

When a new bead appears in the system at $t=J\Delta t$, matrix $\hat{A}_J$ changes to $\hat{A}_{J+1}$, so dynamical equations become diagonal in a new coordinate system. To switch from the old Rouse frame to the new one, we apply
\begin{equation}
    \vec{z} := \hat{P}^{-1}_{J + 1}\vec{x} = \hat{P}^{-1}_{J + 1}\hat{P}_{J}\vec{y} := \hat{T}_{J} \vec{y}. \label{eq: switch0}
\end{equation}
Next, using Eqs. (\ref{eq:propagate0}) and (\ref{eq: switch0}) we relate the joint distributions $\rho_{J}(\vec{y},t)$ and $\rho_{J+1}(\vec{z},t)$ in Rouse frames corresponding to $J$-th and $J+1$-th time intervals respectively as
%\begin{equation}
%    \rho_{J+1}(\vec{z}, t) = \int d\vec{z}_0 \ \rho_{J+1}(\vec{z}_0, J\Delta t)\cdot  {\cal G}_{J+1}(\vec{z}, t-J\Delta t|\   \vec{z}_0), \label{eq:propagate0}
%\end{equation}
\begin{equation}
\begin{aligned}\label{eq:propagate01}
    \rho_{J+1}(\vec{z}_0, J \Delta t) = \int d\vec{y}_0 \ \rho_{J}(\vec{y}_0, (J-1)\Delta t) \\ \times {\cal G}_{J}(\hat{T}_J^{-1}\vec{z}_0, \Delta t|\  \vec{y}_0).
\end{aligned}
\end{equation}
%\begin{equation}
%    \rho_{J+1}(\vec{z}, t) = \int d\vec{y} \ \rho_{J}(\vec{y}, J\Delta t)\cdot  {\cal G}_{J+1}(\vec{z}, t-J\Delta t|\  \hat{T}_{J} \vec{y}), \label{eq:propagate0}
%\end{equation}
%where $t\in [J\Delta t,(J+1)\Delta t]$.
Since the propagator ${\cal G}_{J}(\vec{y}, t|\ \vec{y}_0)$ is Gaussian and $\rho_1(\vec{y}, 0)=\delta(\vec y)$ by assumption, it is easy to see that the overall statistics is going to be zero-mean Gaussian with the covariance matrix determined by the pair correlation function $F_{n,m}(t)$.
%This results in a formula relating $F_{n,m}(t)$ and $F_{n,m}(t + \Delta t)$.
By continuing to perform (\ref{eq: switch0}) and (\ref{eq:propagate01}) every time a new bead appears, we obtain an iterative procedure, which allows us to calculate the exact $F_{n,m}(t)$. %for an arbitrary time dependence of extrusion velocity.
The technical details, which are omitted here for the sake of brevity, can be found in Appendix \ref{appendix:2}.

\section{Continuous limit: Green function approach}\label{sec: continuum}

\begin{figure*}[t!]
    \includegraphics[width=\linewidth]{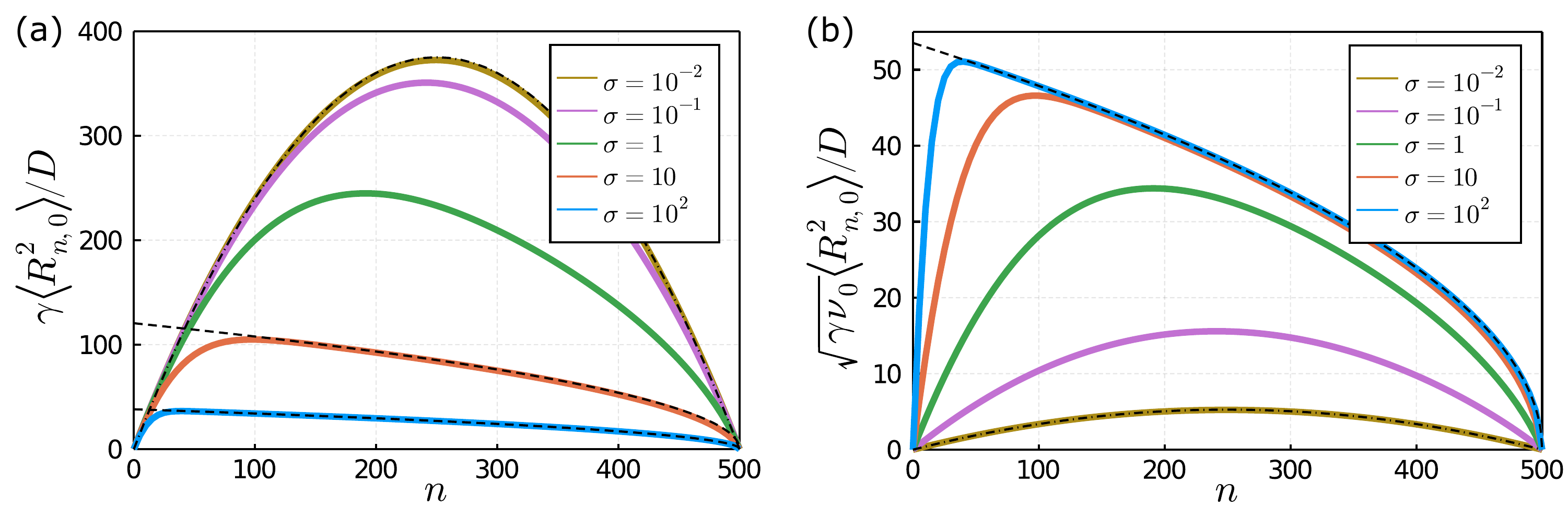}
    \caption{(a) Mean squared separation $\langle R^2_{n,0}\rangle$ between the loop base and the $n$-th bead for the extruded loop of length $N=500$, compared with asymptotic predictions in the near-equilibrium limit ($\sigma\ll 1$, see Eq. (\ref{msd_in_eq})) and in the highly non-equilibrium case ($\sigma\gg 1$ see Eq.~(\ref{envelope})). (b) The same data is shown in different coordinates to reveal the asymptotic behaviour (\ref{envelope}) of $\langle R^2_{n,0}\rangle$ in the limit $\sigma \gg 1$.} \label{Fig: 2}
\end{figure*}

The discrete approach above is general but computationally demanding for large loops. So, as an alternative, we consider the continuum formulation of the Rouse model (see, e.g., Ref. \cite{GKh_1994}), which is justified for sufficiently long polymer segments composed of large number of beads. Indeed, when $N(t)\gg1$, the label of the bead in Eq.~(\ref{"eq:1"}) can be treated as a continuous variable.
Then the position $\vec r(n,t)$ of the $n$-th bead in the loop evolves accordingly to the stochastically forced diffusion equation
\begin{equation}
\frac{\partial \vec r(n,t)}{\partial t}=\frac{k}{\zeta} \frac{\partial^2\vec r(n,t)}{\partial n^2}+ \frac{1}{\zeta}\vec \xi(n,t), \label{continuous}
\end{equation}
which should be supplemented by the zero  conditions $\vec r(0,t)=\vec r(N(t), t)=0$ at the boundaries of the domain $n\in [0,N(t)]$ with $N(t)=\nu_0t$. The random force field in the right hand side of Eq.~(\ref{continuous}) is characterised by zero mean value $\langle \xi_\alpha(n,t)\rangle=0$ and the pair correlator
\begin{equation}
\label{noise_continuum}
\langle \xi_\alpha(n,t_1)\xi_\beta(m,t_2)\rangle = 2\zeta k_BT\delta_{\alpha\beta}\delta(n-m)\delta(t_2-t_1).
\end{equation}
Compared with expression~(\ref{eq:noise_vp}), we have replaced the Kronecker delta symbol $\delta_{nm}$ with the Dirac delta function $\delta(n-m)$.

The exact solution of Eq.~(\ref{continuous}) for a given realization of the noise can be written as
\begin{equation}\label{eq:rnt}
\vec r(n,t)=\frac{1}{\zeta}\int\limits_0^{t}dt_0 \int\limits_{0}^{N(t_0)}dn_0 G(n,t;n_0,t_0)\vec \xi(n_0,t_0),
\end{equation}
where $G(n,t;n_0,t_0)$ represents the Green function of the diffusion equation in a linearly growing domain with zero boundary conditions, which is given by (see Ref.~\cite{Redner_2015}) %and Appendix~\ref{app:GF})
%\begin{widetext}
\begin{equation}
\label{Green_function}
\begin{aligned}
G(n,t,n_0,t_0)=\frac{2 \exp\left[-\frac{\nu_0}{4\gamma} \left(\frac{n^2}{N(t)}-\frac{n_0^2}{N(t_0)} \right)\right]}{\sqrt{N(t_0)N(t)}} \times \\ \sum_{j=1}^{\infty}\sin\left(\frac{j\pi n}{N(t)}\right)\sin\left(\frac{j\pi n_0}{N(t_0)}\right) \exp\left[-\frac{j^2\pi^2\gamma (t-t_0)}{N(t_0)N(t)}\right].
\end{aligned}
\end{equation}
%\end{widetext}
However, this expression is not convenient for subsequent numerical analysis. Instead, we found that it makes sense to use the Poisson summation formula to obtain an alternative expression that is more suitable for numerical evaluation. The details can be found in Appendix~\ref{app:GF}.

Next, substituting Eq.~(\ref{eq:rnt}) into Eq.~(\ref{pair_correlator}) and averaging over noise statistics determined by Eq.~(\ref{noise_continuum}) yields the following integral expression for the pair correlation function of beads coordinates \begin{equation}
\begin{aligned}\label{correlator_continuum}
F_{n,m}(t)= 6D \int\limits_{0}^t dt_0 \mkern-9mu \int\limits_{0}^{N(t_0)} \mkern-9mu dn_0 G(n_1,t;n_0,t_0)G(n_2,t;n_0,t_0).
\end{aligned}
\end{equation}
Also, from Eqs.~(\ref{gyration_radii_def}) and (\ref{correlator_continuum}) one obtains
%\begin{equation}\label{eq:Rgyr_pf2}
%\begin{aligned}
%R_g^2=\frac{6D}{N(t)}\left[\int\limits_{0}^{N(t)}dl\int\limits_{0}^tdt_0\int\limits_{0}^{N(t_0)}dn_0 G^2(n,t;n_0,t_0) - \frac{1}{N(t)}\\ \int\limits_{0}^{N(t)}dn_1\int\limits_{0}^{N(t)}dn_2\int\limits_{0}^tdt_0 \int\limits_{0}^{N(t_0)}dn_0 G(n_1,t;n_0,t_0)G(n_2,t;n_0,t_0)\right]
%\end{aligned}
%\end{equation}
Eq.~(\ref{gyration_app}) in Appendix~\ref{app:GF} for the gyration radius. The remaining series of multiple integrals can be effectively evaluated numerically.

To conclude this section, let us note that Eqs. (\ref{eq:rnt}) and (\ref{Green_function}) suggest the following form of the pair correlation function  and gyration radius: $F_{n,m}(t) = {D}\cdot{\Theta}\left(\frac{n}{N(t)}, \frac{m}{N(t)}, \sigma(t)\right)/\nu_0$ and $R_g^2(t) = D\cdot \Phi(\sigma(t))/\nu_0$, where $\Theta$ and $\Phi$ are some dimensionless functions.
In order to arrive at  these results one should pass to the dimensionless variables  in the expressions for the pair correlation function and gyration radius.

%pay attention to the following property of our model. Consider an arbitrary moment in time $t$ when the loop consists of $N=\nu_0 t$ beads. We can measure the coordinate of the beads and the time in new units, $\tilde n = n/N$ and $\tilde \tau = \nu_0 \tau/N$, which range from $0$ to $1$. If we now turn to these new variables in the expressions for the pair correlation function (\ref{correlator_continuum}) and gyration radius (\ref{gyration_app}), we can get that $\nu_0 F_{n,m}(t)/D = f(\tilde n, \tilde m, \sigma(t))$ and $\nu_0 R_g^2(t)/D = g(\sigma(t))$, where $f??$ and $g?$ are some functions.
%Thus, all the dependence on the parameters of the problem reduces to a single parameter $\sigma$. This means that the degree of non-equilibrium $\sigma$ completely determines the properties of the extruded loop. %and the dependence on other parameters of the problem reduces to a simple rescaling.}
%In particular, two loops with the same values of $\sigma$ may have different values for the gyration radii, but the values of $\nu_0 R_g^2/D$ must be the same.}

\begin{figure*}[t]
    \centering
    \includegraphics[width=\linewidth]{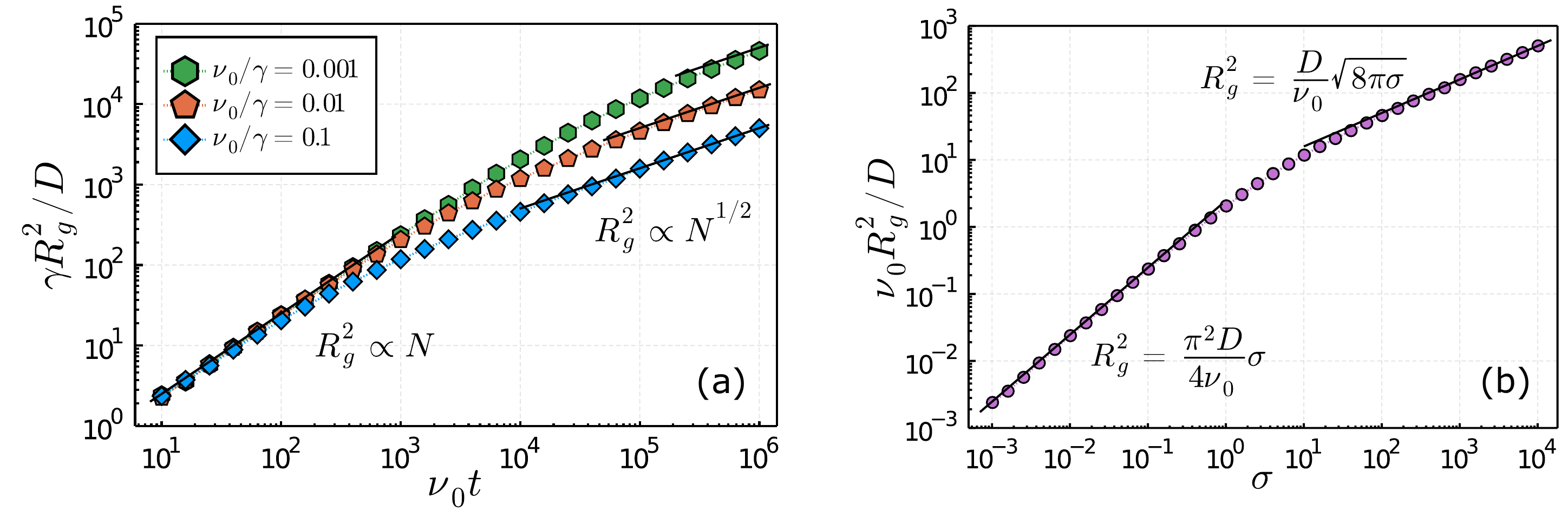}
    \caption{(a) The gyration radii of loops formed by LEFs with different extrusion rates $\nu_0$ depending on their length $N(t)=\nu_0 t$. Clearly, the loops formed by LEFs with larger values of $\nu_0$ turn out to be more compact. (b) The same data, but shown in different coordinates and compared with expressions (\ref{rgyreq}) and (\ref{rgyr}). %The coincidence of all curves underlines the universal role that the parameter $\sigma$ plays in describing the properties of loops.
    }
    \label{Fig: 4}
\end{figure*}

\section{Results and discussion}\label{sec: results}

Confronting the predictions of continuous and discrete models described in Sections \ref{sec: discrete} and \ref{sec: continuum}, respectively, we found that for a moderately large loop length $N$ two semi-analytical approaches match each other nearly perfectly.
The only difference appears close to the right boundary (where the loop is getting extruded), but this local discrepancy is not relevant for loop averaged metrics, so we have managed to obtain consistent predictions for the gyration radii and the  contact frequency enhancement (see below) using both approaches.
Given this agreement, only the data extracted from the continuous model are shown in the plots throughout the rest of the paper.

%\subsection{Mean squared displacement.}

%\begin{figure}[h]
%    \centering
%    \includegraphics[width=0.5\textwidth]{beads.pdf}
%    \caption{$\langle r_n^2\rangle$ as a function of elapsed time $\Delta t$ compared with (\ref{equil}) and (\ref{envelope}), assuming that $\gamma/\nu_0=10$.}
%    \label{Fig: 3}
%\end{figure}

\subsection{Mean squared separation}
\label{sec: mss}

We start presentation of results with Fig.~\ref{Fig: 2} which demonstrates the mean squared separation $\langle R_{n,0}^2(t)\rangle=\langle (\vec r_n(t)-\vec r_0(t))^2\rangle$ between loop base and the bead inside a loop as a function of the bead number $n$ for a loops of the  contour length $N=500$.
Different curves corresponds to the different  values of the non-equilibrium degree $\sigma$ (see Eq.~(\ref{eq: sigma})).
Here and in what follows we assume that parameters $D$ and $\gamma$ associated with the physical properties of the polymer chain are fixed so that  $\sigma$ is varied by changing the extrusion velocity $\nu_0$.
Fig.~\ref{Fig: 2}a tells us that at $\sigma\ll 1$ the shape of the curve $\langle R^2_{n,0}\rangle$ is indistinguishable from the  equilibrium profile (see Appendix~\ref{sec: app_equilibrium })
\begin{equation}
\label{msd_in_eq}
\langle R_{n,0}^2(t)\rangle_{\text{eq}} =\frac{3D}{\gamma}\frac{n(N(t)-n)}{N(t)}.
\end{equation}
However, as $\sigma$ is getting larger, the curve $\langle R^2_{n,0}\rangle$ becomes more and more asymmetric, and at $\sigma\gg 1$ the numerical fit revealed the following asymptotic behaviour
\begin{equation}
\langle R^2_{n,0}(t)\rangle\approx 3\sqrt{\frac{2}{\pi}}\frac{D\sqrt{N(t)-n}}{\sqrt{\gamma\nu_0}}, \label{envelope}
\end{equation}
which is valid for $n\gg\sqrt{\gamma t}$, see Fig.~\ref{Fig: 2}b.
In Section \ref{sec: discussion}  we will explain how to derive Eq. (\ref{envelope}) analytically.

%It also demonstrates that highly non-equilibrium regimes have scale-free properties - both diagonal and off-diagonal correlators are described by universal dependencies.

\subsection{Radius of gyration}

From Fig.~\ref{Fig: 2}a we may conclude that a non-equilibrium loop composed of $N$ beads is more compact than its equilibrium counterpart of the same contour length. To quantify this difference we next plot in Fig.~\ref{Fig: 4}a the gyration radius $R_g^2$ as a function of the number of beards $N=\nu_0 t$ in the growing loop for  different extrusion rates $\nu_0$.

%of the time $t$ since the attachment of the LEF or the number of beards $N=\nu_0 t$ in the loop.
%the non-equilibrium degree $\sigma$ (or, equivalently, of the time since the attachment of the LEF, as the parameter $\sigma$ is proportional to the number of beards $N=\nu_0 t$ in the loop, see Eq. (\ref{eq: sigma})).

As discussed in Section \ref{sec: model}, when the loop grows, it gradually becomes more and more non-equilibrium, which is clearly seen from Fig.~\ref{Fig: 4}a. Indeed, the initial quasi-equilibrium stage of loop evolution is characterised by the usual linear proportionality between the gyration radius and loop size ($R_g^2\propto N(t)$ at $\sigma\lesssim1$), whereas the further non-equilibrium stage establishes the square root scaling law ($R_g^2\propto \sqrt{N(t)}$ for $\sigma\gg 1$).
%and the loops formed by LEFs with larger values of $\nu_0$ turn out to be more compact.
To emphasize the crucial role of the  parameter $\sigma$ when describing the properties of loops, we present the data shown in Fig.~\ref{Fig: 4}a in new coordinates. Now the $Y$-axis corresponds to $\nu_0 R_g^2/D$ and the $X$-axis -- to the values of $\sigma$, see Fig.~\ref{Fig: 4}b. All data points fall on the universal curve in agreement with the general arguments  presented at the end of Section~\ref{sec: continuum}.

\begin{figure*}[!t]
    \centering
    \includegraphics[width=\linewidth]{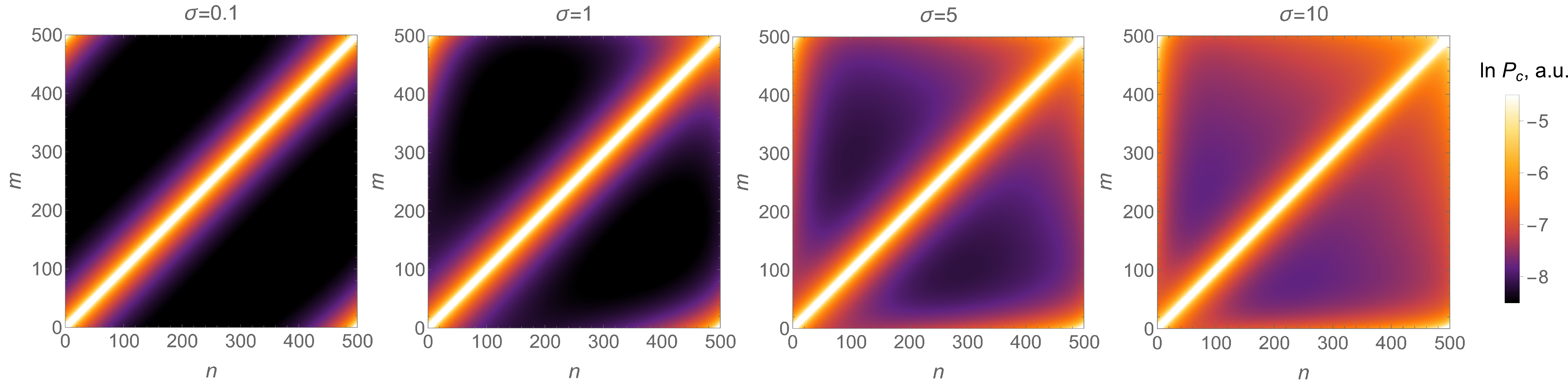}
    \caption{Contact maps for four loops of the same length $N=500$ which differ by the value of parameter $\sigma$ (or equivalently, by the extrusion velocities of LEFs in their bases). The color represents the logarithm of the contact probability $P_c(n,m;t)$ defined in equation (\ref{eq: contact}). These maps were generated semi-analytically using Eqs.~(\ref{Green_function}) and (\ref{correlator_continuum}).The resolution of maps in each direction is equal to $\Delta n = \Delta m = 2$.}
    \label{Fig: 5}
\end{figure*}

Beyond the proportionality dependencies, the quasi-equilibrium radius of gyration is given by (see Ref.~\cite{GKh_1994} and Appendix~\ref{sec: app_equilibrium })
\begin{equation}
R_{g,\text{eq}}^2(t)=\frac{D}{4\gamma}N(t) \label{rgyreq},
\end{equation}
while at far-from-equilibrium conditions one finds
\begin{eqnarray}\label{rgyr}
R_g^2(t)\approx 2\sqrt{\frac{2}{\pi}}D\sqrt{\frac{N(t)}{\gamma\nu_0}}.
\end{eqnarray}
The later expression is obtained from Eqs.~(\ref{gyration_radii_def}) and (\ref{envelope}) under the assumption of negligible correlations between most of the beads (see section \ref{sec: discussion} for justification of this calculation), and it indeed provides a fit to large-$\sigma$ asymptotic behavior of $R_g^2$ as shown in Fig.~\ref{Fig: 4}b.

From Eqs.~(\ref{rgyreq}) and (\ref{rgyr}) we find that the ratio of the true size of the non-equilibrium loop to its naive equilibrium estimate is controlled by the parameter $\sigma$
\begin{eqnarray}
\frac{R_g^2}{R_{g,\text{eq}}^2}=\frac{8\sqrt2}{\pi^{3/2}\sqrt{\sigma}}
\end{eqnarray}
and this ratio is small for $\sigma\gg 1$. In a sense, the more compact conformation of non-equilibrated loops as compared with that of statistically static loops is not unexpected. Small value of $\sigma$ means that the looped segment  has enough time to explore the phase space of possible conformations before its length will be significantly changed due to ongoing extrusion process. By contrast, at large $\sigma$, the overwhelming majority of beads that are brought into proximity in the region near loop base do not have enough time to relax to their joint near-equilibrium statistics dictated by the current loop length. Importantly, this difference cannot be accounted as a simple renormalization of the parameters entering expression for gyration radii of the equilibrium loop. Indeed, Eq.  (\ref{rgyr}) show that non-equilibrium nature of the loop extrusion process entails a different type of scaling behaviour at $\sigma\gg 1$.

\subsection{Contact probability}

\begin{figure}[t]
    \centering
    \includegraphics[width = \linewidth]{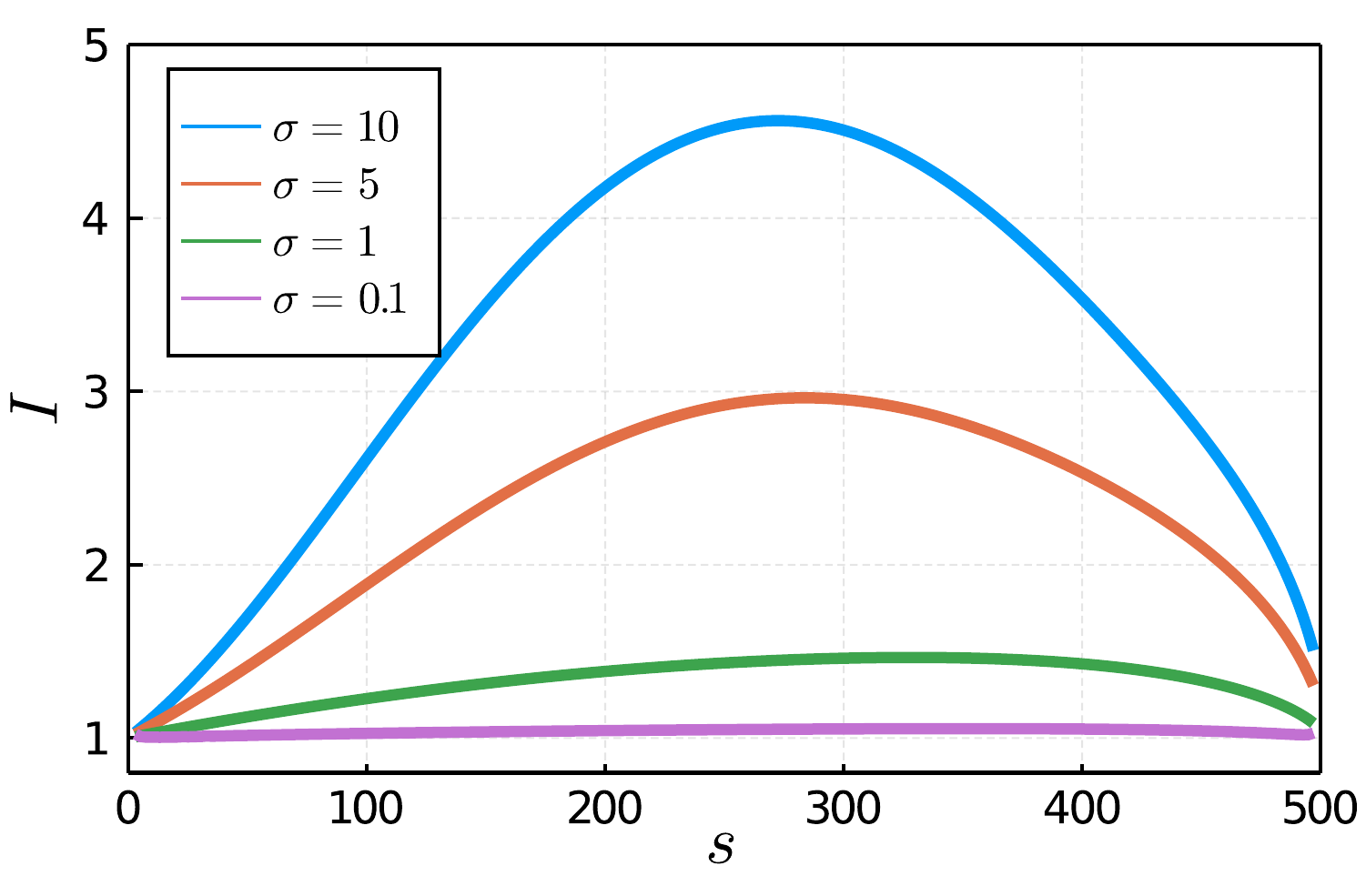}
    \caption{The normalized excess of contacts $I(s)$ (see Eq. (\ref{cp_enhancement})) between beads inside the extruded loop over contacts between the corresponding beads of the equilibrium loop of the same contour length $N=500$. The summation in expression (\ref{eq:sumI}) was performed with a step of 2 beads using the contact maps shown in Fig.~\ref{Fig: 5}.} %Solid lines correspond to the continuous model, and dashed lines - to the discrete model.
    \label{Fig: 6}
\end{figure}

It is natural to suggest that since more non-equilibrium loop occupies smaller volume, than larger value of extrusion velocity must entail higher frequency of inter-beads physical contacts inside the loop. The contact probability maps depicted in Fig.~\ref{Fig: 5} clearly confirm these expectations.
To  quantify the  increase in contact frequency between monomers on the non-equilibrium loop, we introduce the following metric
\begin{equation}
\label{cp_enhancement}
I=\frac{P_c(s;t)}{P_c^{eq}(s)},
%=\frac{\langle R_{n,n+s}^2\rangle_{eq}^{3/2}}{N(t)-s + 1}\sum_{n=0}^{N(t)-s}\frac{1}{\langle R_{n,n+s}^2(t)\rangle^{3/2}}.
\end{equation}
where
\begin{equation}\label{eq:sumI}
P_c(s;t)=\frac{\sum_{n=0}^{N(t)-s} P_c(n,n+s;t)}{N(t)-s+1},
\end{equation}
is the loop-averaged contact probability. In other words, $P_c(s;t)$ is determined as the averaging of the pairwise contact probability $P_c(n,m;t)$ (see Eq.~(\ref{eq: contact})) over all pairs of beads separated by a given contour distance $s$. The corresponding equilibrium value $P_{c}^{\text{eq}}$ entering Eq.~(\ref{cp_enhancement}) is given by Eq.~(\ref{cp_eq}). Fig.~\ref{Fig: 6} indicates that maximal (relative) enhancement of interactions is observed for pairs of beads separated by the contour distance about the half loop size.

%\begin{equation}
%I=\frac{P_c(s;t)}{P_c^{eq}(s)}=\frac{\langle R_{n,n+s}^2\rangle_{eq}^{3/2}}{N(t)-s + 1}\sum_{n=0}^{N(t)-s}\frac{1}{\langle R_{n,n+s}^2(t)\rangle^{3/2}},
%=3\sqrt{3}\left( \frac{D}{\gamma}\right)^{3/2}\frac{s^{3/2}\sqrt{N(t)-s}}{N(t)^{3/2}}\sum_{n=0}^{N(t)-s}\frac{1}{\left(\langle \vec r^2(n,t)\rangle+\langle \vec r^2(n+s,t)\rangle-2 \langle \vec r(n,t)\cdot\vec r(n+s,t)\rangle\right)^{3/2}},
%\end{equation}
%where $1\le s\le N(t)-1$.
%Obviously
%\begin{eqnarray}
%&&\langle R_{n,n+s}^2(t)\rangle= \langle \vec r^2(n,t)\rangle+\langle \vec r^2(n+s,t)\rangle-2 \langle \vec r(n,t)\cdot\vec r(n+s,t)\rangle,\\
%&& \langle R_{n,n+s}^2\rangle_{eq}=\frac{3D}{\gamma}\frac{s(N-s)}{N}.
%\end{eqnarray}

%As $\sigma$ grows, contact probabilities between beads also increase (see Figs \ref{Fig: 5} and \ref{Fig: 6}). Here, we compare the average increase in contact frequency between beads at contour distance $s$ from each other between non-equilibrium and equilibrium regimes as

%where the second equality follows from (\ref{eq: contact}) in the Appendix.
%For large values of $s$, a formal discrepancy between discrete and continuous approaches is observed, which is for reasons discussed in the beginning of this section.

%, both for $\langle r^2(n)\rangle$ and $R_g^2$ (see Fig. \ref{Fig: 4}). After a significant amount of time has elapsed, and the growing boundary gets far away from a selected bead, we observe a gradual relaxation towards its equilibrium expectation value (while the loop as a whole, of course, continues to be highly non-equilibrium).

\subsection{Analytical solution in the limit $\sigma\gg1$}
\label{sec: discussion}

Surprising simplicity of Eq.~(\ref{envelope}) guessed to fit the large-$\sigma$ behaviour of the MSD and gyration radius predicted by our semi-analytical computational schemes calls for its analytical derivation.
Here we  provide such a derivation using an approximate (but asymptotically correct) solution of Eq. (\ref{continuous}) in the limit of strongly non-equilibrium loop.

Comparing different terms in Eq.~(\ref{continuous}), we conclude that the beads whose dynamics is strongly affected by the zero condition at the left boundary of the interval $n\in [0, N(t)]$ are those with the label $n\ll \sqrt{\gamma t}$. According to Eq.~(\ref{eq: sigma}), large value of the parameter $\sigma$ is equivalent to the inequality $\sqrt{\gamma t}\ll N(t)$. Therefore, majority of beads  inside the growing loop, which is characterised by $\sigma\gg 1$, do not feel the presence of the  boundary condition at $n=0$. This allows us to pass to the simplified problem defined at the semi axis. More specifically, we ignore the left boundary and pass to the new variable $l=-n+\nu_0t$ in Eq.~(\ref{continuous}), which corresponds to the bead number measured from the right end of the loop. Then, Eq.~(\ref{continuous}) yields
\begin{equation}
\frac{\partial \vec r(l,t)}{\partial t}=\gamma \frac{\partial^2\vec r(l,t)}{\partial l^2}-\nu_0\frac{\partial \vec r(l,t)}{\partial l}+ \frac{1}{\zeta}\vec \xi(\nu_0t-l,t),
\end{equation}
where $l\ge 0$ and $\vec r(0,t)=0$.  The solution to this problem with zero initial condition is given by
\begin{equation}
    \vec r(l,t)=\frac{1}{\zeta}\int\limits_0^{t}dt_0 \int\limits_{0}^{\infty}dl_0 g(l,t;l_0,t_0)\vec \xi(\nu_0t_0 - l_0,t_0),
\end{equation}
where
\begin{equation}
\begin{aligned}\label{eq: green_simplified}
g(l,t;l_0,t_0)=\frac{\exp\left[\frac{\nu_0}{2\gamma}(l-l_0)-\frac{\nu_0^2}{4\gamma}(t-t_0)\right]}{\sqrt{4\pi\gamma(t-t_0)}} \times\\
\left(\exp\left[-\frac{(l-l_0)^2}{4\gamma (t-t_0)}\right]-\exp\left[-\frac{(l+l_0)^2}{4\gamma (t-t_0)} \right]\right)
\end{aligned}
\end{equation}
%\begin{eqnarray}
%&&g(l,t;l_0,t_0)=%\frac{2}{\pi}\exp\left(\frac{\nu_0(l-l_0)}{2\gamma}-\frac{\nu_0^2(t-t_0)}{4\gamma}\right)\int_{0}^{\infty}dq\sin\left(ql\right)\sin\left(ql_0\right)
%\exp\left(-q^2\gamma (t-t_0)\right)= \label{green:simplified}\\
%\frac{\exp\left[\frac{\nu_0}{2\gamma}(l-l_0)-\frac{\nu_0^2}{4\gamma}(t-t_0)\right]}{\sqrt{4\pi\gamma(t-t_0)}}\\
%&&\times\left(\exp\left[-\frac{(l-l_0)^2}{4\gamma (t-t_0)}\right]-\exp\left[-\frac{(l+l_0)^2}{4\gamma (t-t_0)} \right]\right),
%\end{eqnarray}
is the Green function of the drift-diffusion equation with zero boundary condition at the edge of positive semi-axis. As was established in section \ref{sec: mss},  at  $\sigma\gg1$ in the region $n\gg \sqrt{\gamma t}$ the loop is characterised by the universal profile of $\langle R_{n,0}^2\rangle$ (see Eq.~(\ref{envelope}) and Fig.~\ref{Fig: 2}b), which is a function only of $N(t)-n = l$. In other words, it is independent of time in terms of $(l,t)$ variables. Thus, we expect to obtain the correct asymptotic behavior by taking the limit $t \to +\infty$, which is going to make the result a function of $l$ exclusively. So, after averaging over noise statistics, we arrive at (see Appendix \ref{app: analytics})
\begin{equation}
\begin{aligned}\label{eq: mss_approx}
    \langle r^2(l)\rangle = \lim_{t\to +\infty} 6D\int\limits_0^tdt_0\int\limits_0^{+\infty}dl_0g^2(l,t;l_0,t_0) \approx \\ \approx3\sqrt{\frac{2}{\pi}}\frac{D\sqrt{l}}{\sqrt{\gamma\nu_0}}.
\end{aligned}
\end{equation}
Clearly, $\langle R_{n,0}^2(t)\rangle=\langle r^2(N(t)-n)\rangle$ and thus Eq. (\ref{eq: mss_approx}) yields Eq. (\ref{envelope}).

%which reproduces the universal numerical fit that we observed earlier.
Similarly, we can address the problem of calculating the pair correlator $\langle(\vec{r}(l_1,t) \cdot \vec{r} (l_2,t)\rangle$.
%It is, both from physical considerations and semi-analytical results, apparent that in highly non-equilibrium regimes beads would be in contact only with a relatively small fraction of the loop.
To carry out this calculation we can introduce $l_1 = l$ and $l_2 = l + \Delta l$, and use relative correlation distance $ \Delta l/l$ as a small parameter. By performing steps analogous to the derivation of Eq. (\ref{eq: mss_approx}), but keeping terms that are up to $O((\Delta l/l)^2)$ in binomial expansions, we arrive at the following leading order asymptotic expression (see Appendix \ref{app: analytics})
\begin{equation}
\label{pair_corr_approx}
\begin{aligned}
    \langle \vec{r}(l + \Delta l, t) \cdot \vec{r}(l, t) \rangle \approx 3\sqrt{\frac{2}{\pi}}\frac{D\sqrt{l}}{\sqrt{\gamma\nu_0}} \cdot \exp \left(-\frac{\nu_0}{8 \gamma} \frac{(\Delta l)^2}{l} \right) \\ - \frac{3D}{2 \gamma}|\Delta l| \cdot \text{Erfc}\left( \sqrt{-\frac{\nu_0}{8 \gamma} \frac{(\Delta l)^2}{l}}\right).
\end{aligned}
\end{equation}
This result is self-consistent
% with the approach that was employed to derive it,
 since it demonstrates that, indeed, for $l\gg \gamma/\nu_0$ (clearly, $\gamma/\nu_0\ll N(t)$ at $\sigma\gg1$) the relative correlation length is
\begin{equation}
    \frac{\Delta l}{l} \sim \sqrt{\frac{\gamma}{l\nu_0}} \ll 1.
\end{equation}
Thus, it serves as a justification for the assumption of negligible correlations between most of the beads, which we used earlier to obtain Eq. (\ref{rgyr}).

\section{Conclusion and outlook}

\begin{figure}[t]
    \centering
    \includegraphics[width = \linewidth]{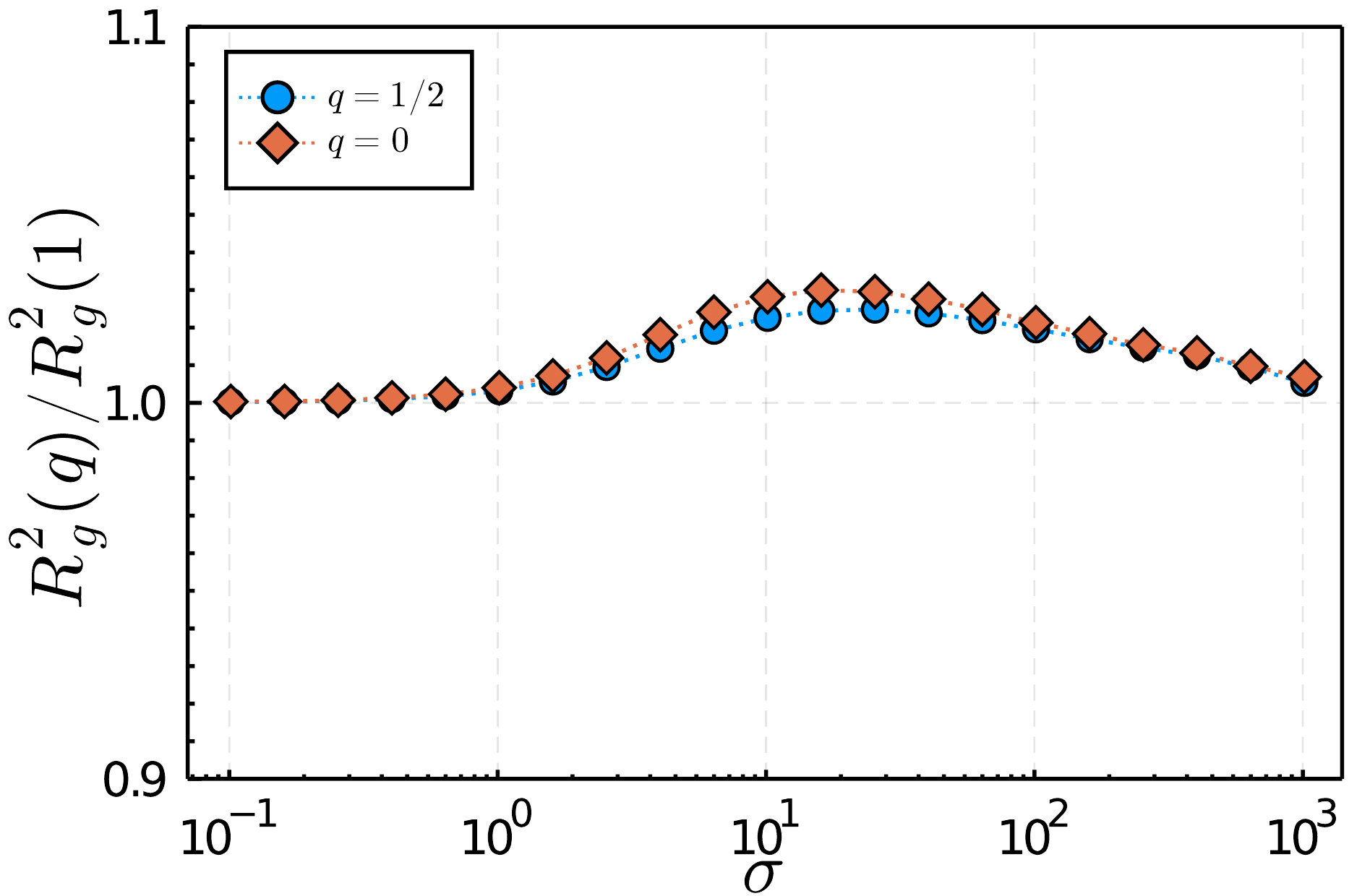}
    \caption{Ratio of  gyration radius of the non-equilibrium loop extruded by the LEF with symmetry score $q$ (see Ref.~\cite{Golfier_2020}) to that of the loop  generated via purely one-sided extrusion (i.e. $q=1$), as a function of $\sigma$. Note that $q = 0$ corresponds to the perfectly symmetric two-sided extrusion, while the choice $q=1/2$ means that the LEF extrudes the right arm of the loop $3$ times faster that the left arm.}
    \label{Fig: 7}
\end{figure}

To summarize, we explored theoretically the conformational statistics  of growing loops of  ideal polymer chain. Our analysis demonstrated that statistical properties of an extruded loop are determined by the dimensionless parameter $\sigma$ defined as a ratio of the loop relaxation time and the time required to extrude this loop. When the parameter of non-equilibrium is small, $\sigma\ll 1$, the loop approaches the equilibrium coil in its statistical properties, which is reflected in the linear scaling of the gyration radius with the loop length. In the opposite case, when $\sigma\gg 1$, the highly non-equilibrium nature of the loop  manifests itself in increased contact frequencies between monomers inside the looped region and the square root dependence of the gyration radius on the loop length. These results are in accord with the recent numerical studies reported that faster extrusion produces more compact loops and more bright contact maps \cite{Nuebler_2018}.

Thus far, we have assumed that the LEF extrudes polymer chain from one side. While the first experimental demonstration of the loop extrusion reported that yeast condensins extrude DNA loops in almost purely asymmetric (one-sided) manner \cite{Ganji_2018}, subsequent single-molecule experiments showed that human condensins may exhibit both one-sided and two-sided loop extrusion activity \cite{Golfier_2020,Kong_2020}. Besides, DNA loop extrusion by another SMC complex -- cohesin -- is found to be largely symmetric \cite{Golfier_2020,Davidson_2019, Kim_2019}. The details of in vivo loop extrusion remain to be unknown, since all above mentioned results are obtained in in-vitro conditions. However theoretical modelling indicates that an assumption of pure one-sided loop extrusion cannot explain some important chromosome organization phenomena in living cells \cite{Banigan_2019,Banigan_2020a,Banigan_2020b}.

How does incorporation of two-sided extrusion modify the conclusions of the above analysis? Direct generalization of our approach to the case of two-sided extrusion (see Appendix~\ref{app:TS} for details) demonstrates that all of the aforementioned predictions retain their asymptotic form. In particular, Fig.~\ref{Fig: 7} shows that the size of loops produced by two-sided LEFs is larger that of the loops generated via one-sided extrusion, but the magnitude of this effect does not exceed several percents. Thus, from the perspective of single-loop statistics adopted here, one-sided and two-sided extrusion models are practically indistinguishable.

Much remains to be done on the side of analytical theory. Further research beyond the one-loop level should illuminate how a dynamic array of growing, colliding and disappearing loops generated by the loop extrusion factors that exchange between polymer and solvent affects the conformational properties of a Rouse chain. Also, we expect that confronting analytical predictions with experimental data may reveal the necessity for more sophisticated polymer models incorporating excluded volume repulsion, hydrodynamic interaction and bending rigidity. %The investigations in these directions have already began.

%Discuss the case of time-dependent extrusion velocity.
%The extrusion velocity depends on the DNA stretching which grows with time.
%Explain that the extrusion velocity depend on the time elapsed since the start of extrusion process.
%To the best of our knowledge, there is no closed-form analytical expression of the Green function of Eq. () in the case of non-linearly growing domain.
%Figure 7: gyration radii of the loop having size $N=500$ and $\sigma=50$ for different time dependencies of the extrusion velocity: (a) $\nu(t)=const=\nu_0$,  and  $v(t)=At^{\alpha}$ with (b) $\alpha=-1/3$ (c) $\alpha=-1/2$, (d) $\alpha=-2/3$.

%Estimate $\sigma$ for typical \textit{in vivo} conditions: $v_0=1kbp/sec$, $\lambda=100kbp$.
%Stress  high viscosity of the intracellular environment.

%$k=\frac{3k_BT}{a^2}$ - entropic elasticity

%$\zeta=6\pi\eta a$ - friction coefficient

%$D=\frac{k_BT}{\zeta}$ - diffusivity of a bead

%We assumed that loop base is fixed in space.
%How does this affect the model predictions?
%Explain that results will be the same in the limits $\sigma\ll 1$ and $\sigma\gg 1$.

%Assumptions: we neglected the forces arising from the hydrodynamic self-interaction of the polymer, no excluded volume effects.

%Future directions: $\beta$-polymer model with loop extrusion, ensemble of loops.

\section*{Data AVAILABILITY}

The data that support the findings of this study are available from the corresponding author upon reasonable request.

\begin{acknowledgments}
The authors thank Vladimir Lebedev and Igor Kolokolov for helpful discussion. This work was supported by Russian Science Foundation, Grant No. 20-72-00170.
\end{acknowledgments}

\appendix

\begin{widetext}

\section{Basic properties of an equilibrium loop} \label{sec: app_equilibrium }

The probability distribution of the separation vector $ \vec R_{n,m}(t)=\vec r_n(t)-\vec r_m(t)$ between $n$-th and $m$-th beads of an equilibrium loop having size $N$ is given by (see Ref.~\cite{GKh_1994})
\begin{equation}
P_{\text{eq}}(\vec R_{n,m})=\left(\frac{N\gamma}{2\pi Ds(N-s)}\right)^{3/2}\exp\left(-\frac{N\gamma R_{n,m}^2}{2 Ds(N-s)}\right), \label{equil}
\end{equation}
where $s=|n-m|$.
From Eq.~(\ref{equil}) one obtains the following results for the mean squared (physical) separation between two beads
\begin{equation}
\langle R_{n,m}^2\rangle_{\text{eq}} = \langle (\vec r_n-\vec r_m)^2\rangle=\int P_{\text{eq}}(\vec R_{n,m}) R_{n,m}^2d^3R_{n,m}=\frac{3D}{\gamma}\frac{s(N-s)}{N},
\end{equation}
for the radius of gyration
\begin{equation}
R_{g,\text{eq}}^2=\frac{1}{2N^2}\sum_{n,m=0}^{N-1}\langle(\vec r_{n}(t)-\vec r_m(t))^2\rangle=\frac{1}{2N^2}\int_0^N\int_0^N dndm \langle R_{n,m}^2\rangle_{\text{eq}} =\frac{D}{4\gamma}N \label{rgyreq_app},
\end{equation}
and for the probability of contact of the pair of beads separated by the (contour) distance $s$
\begin{equation}
\label{cp_eq}
P_{c}^{\text{eq}}(s)=%\sqrt{\frac{6}{\pi}}\frac{a_0^3}{\langle \vec R^2(n)\rangle^{3/2}}=
\text{Prob}[R_{n,m}<a_0]\approx\frac{4}{3}\pi a_0^{3}\int P_{\text{eq}}(\vec R_{n,m})\delta(\vec R_{n,m})d^3R_{n,m}=\frac{4}{3}\pi a_0^{3}P_{\text{eq}}(0)=
\sqrt{\frac{6}{\pi}}\left(\frac{N\gamma}{3D}\right)^{3/2}\frac{a_0^3}{[s(N-s)]^{3/2}}.
\end{equation}

\section{Discrete model}\label{appendix:2}

The propagator of  Eq. (\ref{"eq:FP0"}) is given by
\begin{equation}
    {\cal Q}_j(y_j, t| y_{j0})=\frac{1}{\sqrt{2 \pi D \sigma_j (t)}}  \exp\left(-\frac{[y - \mu_j(t)]^2}{2 D \sigma_j(t)}\right), \label{eq:ind}
\end{equation}
where
\begin{eqnarray}
&&\sigma_j(t) =  - \frac{1}{\lambda_j} (1 - e^{-2|\lambda_j| t}),\\
&&\mu_j(t) = y_{j,0} \cdot e^{-|\lambda_j| t},\\
&&\lambda_j = -2 \gamma \cdot (1 + \cos{\frac{\pi j}{J}}),
\end{eqnarray}
and $j = 1, 2 \dots J - 1$.

As was explained in the main text, the probability density $\rho_{J}(\vec{y}, t)$ is the zero mean normal distribution.
We then substitute into Eq. (\ref{eq:propagate0}) the Gaussian ansatz
\begin{equation}
    \rho_J(\vec{y}, t) \propto \exp\left( -\frac{\vb{y}^T \cdot \hat{R}^{-1}_J (t) \cdot \vb{y}}{2D}\right),
\end{equation}
and
\begin{equation}
    \rho_{J+1}(\vec{z}, t) \propto \exp\left( -\frac{\vb{z}^T \cdot \hat{R}^{-1}_{J+1} (t) \cdot \vb{z}}{2D}\right) \label{eq: rnew},
\end{equation}
where $\hat{R}_J (t)$ and $\hat{R}_{J+1} (t)$  are the matrices of covariances in the  Rouse frames corresponding to the time intervals $[(J-1)\Delta t, J\Delta t)$ and $[ J\Delta t,(J+1)\Delta t)$, respectively.
Using the explicit form of the function $ {\cal G}_{J}(\vec{y}, t|\  \vec{y}_0)=\prod_{j = 1}^{J} {\cal Q}_j(y_j, t| y_{j0})$ with ${\cal Q}_j(y_j, t| y_{j0})$ given by Eq. (\ref{eq:ind})
%Due to Gaussian nature of (\ref{eq:ind}) and, subsequently, (\ref{eq:product}), we are able to perform the integration for an arbitrary $M$ in a convenient matrix form. As a result, we arrive at
we perform integration in Eq. (\ref{eq:propagate01}) and find the following  relation
\begin{equation}
    \hat{R}^{-1}_{J+1} (\tau)  = \hat{T}^{T}_J \cdot \hat{\sigma}_J^{-1}(\tau - t) \cdot \hat{T}_J
     - (\hat{M}_J(\tau - t) \cdot \hat{\sigma}_J^{-1} (\tau - t)\cdot \hat{T}_J)^{T} \cdot \hat{K}(t,
    \tau) \cdot (\hat{M}_J(\tau - t) \cdot \hat{\sigma}_J^{-1} (\tau - t)\cdot \hat{T}_J), \label{eq: iterate}
\end{equation}
where
\begin{equation}
\hat{K}(t,\tau) = (\hat{R}_J^{-1}(t) + \hat{M}_J^{T}(\tau - t) \cdot \hat{\sigma}_J^{-1}(\tau-t)\cdot \hat{M}_J(\tau-t))^{-1},
\end{equation}
\begin{equation}
    \hat{T}_{J} = \hat{P}^{-1}_{J+1} \hat{P}_{J}, \quad \quad \hat{P}^{J}_{n,m} \propto \cdot \sin\left(\pi \frac{ n \cdot m}{J} \right), \quad
\end{equation}
\begin{equation}
    \sigma^{J}_{n,m} = \delta_{n,m} \cdot \frac{e^{-2 |\lambda_m|t}-1 }{|\lambda_m|}, \quad \quad M_{nm}^{J}(t) = \delta_{nm} \cdot e^{-|\lambda_m|t},
\end{equation}
and $n, m = 1, 2 \ \dots \ J - 1$.
%Here, the index $J$ refers to the number of beads, and indices $n,m$ - to the matrix elements.
 Equation (\ref{eq: iterate}) can be easily applied in an iterative computational scheme allowing us to calculate the covariance matrix of the Rouse modes at an arbitrary time moment.
To describe the covariance matrix  of the beads' coordinates, as it was introduced in Eq. (\ref{pair_correlator}), one should substitute $\vec{z} = \hat{P}^{-1}_{J + 1} \vec{x}$ into Eq. (\ref{eq: rnew}), perform the matrix inversion, and multiply the result by a factor of 3 to account for dimensionality, i.e.
\begin{eqnarray}
F^{J}_{n,m} (t) = 3 \cdot \left(\hat{P}_{J} \cdot  \hat{R}^{-1}_{J} (t) \cdot \hat{P}^{-1}_{J}\right)^{-1}_{n,m},
\end{eqnarray}
where  $F^{J}_{n,m} (t)$ denotes the pair correlation function $F_{n,m} (t)$ during the time interval $t\in[(J-1)\Delta t, J\Delta t)$.
%In particular, we can put $\tau = t + \Delta t$ to obtain the covariance matrix right at the start of the next time interval, so that

\section{Green function.}\label{app:GF}

The Green function $G(n,t;n_0,t_0)$ of Eq.~(\ref{continuous}) is defined as the solution to equation
\begin{equation}
\frac{\partial G}{\partial t}=\gamma \frac{\partial^2G}{\partial n^2},
\end{equation}
%where $\gamma=k/\zeta$,
with the initial condition $G(n,t_0;n_0,t_0)=\delta(n-n_0)$ and the boundary conditions
$G(0,t;n_0,t_0)=G(N(t),t;n_0,t_0)=0$, where  $0\le n_0\le N(t_0)$, $0\le n\le N(t)$ and $t\ge t_0$.

An exact solution to this initial-boundary-value problem has been constructed in Ref.~\cite{Redner_2015} and it is given by Eq.~(\ref{Green_function}) in the main text.
%\begin{equation}
%\label{Green_function_app}
%G(n,t;n_0,t_0)=\frac{2}{\sqrt{N(t_0)N(t)}}\exp\left(-\frac{\nu_0(n^2-n_0^2)}{4\gamma N(t)}\right)\sum_{j=0}^{\infty}\sin\left(\frac{j\pi n}{N(t)}\right)\sin\left(\frac{j\pi n_0}{N(t_0)}\right)\exp\left(-\frac{j^2\pi^2\gamma (t-t_0)}{N(t_0)N(t)}\right).
%\end{equation}
To make this formula more suitable for numerical evaluation, we apply the Poisson summation formula
\begin{eqnarray}
\label{Poisson}
\sum_{j \in \mathbb{Z}} f(j) = \sum_{m \in \mathbb{Z}} \hat f(m), \quad \hat f(m) = \int_{-\infty}^{\infty} dx e^{-i 2 \pi m x} f(x),
\end{eqnarray}
which allows us to pass from  Eq. (\ref{Green_function}) to the faster converging representation of the Green function
\begin{equation}
\begin{aligned}
G(n,t;n_0,t_0)=\frac{1}{\sqrt{4 \pi \gamma (t-t_0)}}\exp\left[-\frac{(n^2 t_0-n_0^2 t)}{4\gamma t t_0}\right] \times\\ \sum_{m \in \mathbb{Z}} \left[ \exp \left( - \dfrac{(nt_0 - n_0 t+2m \nu_0 t t_0)^2}{4\gamma t t_0 (t-t_0)}\right) - \exp \left( - \dfrac{(nt_0 + n_0 t+2m \nu_0 t t_0)^2}{4\gamma t t_0 (t-t_0)}\right) \right],
\end{aligned}
\end{equation}
which we use to compute the radius of gyration
%\begin{eiven byquation}
%R_g^2=
%.
%\end{equation}
%Gyration radii:
\begin{equation}\label{gyration_app}
R_g^2(t)=%\int\limits_{0}^{N(t)}\int\limits_{0}^{N(t)}\frac{dn_1dn_2}{2N(t)^2}\langle(\vec r(n_1,t)-\vec r(n_2,t))^2\rangle=
\frac{6D}{N(t)}\left[\int\limits_{0}^{N(t)} \mkern-9mu  dn\int\limits_{0}^tdt_0 \mkern-9mu  \int\limits_{0}^{N(t_0)} \mkern-9mu  dn_0 G^2(n,t;n_0,t_0) - \frac{1}{N(t)}\int\limits_{0}^{N(t)} \mkern-9mu  dn_1 \mkern-9mu  \int\limits_{0}^{N(t)} \mkern-9mu dn_2\int\limits_{0}^tdt_0 \mkern-9mu  \int\limits_{0}^{N(t_0)} \mkern-9mu  dn_0 G(n_1,t;n_0,t_0)G(n_2,t;n_0,t_0)\right].
\end{equation}
To derive Eq. (\ref{gyration_app}) one should substitute Eq. (\ref{correlator_continuum}) into the  definition $R_g^2(t)=\frac{1}{2N(t)^2}\int_{0}^{N(t)}\int_{0}^{N(t)}dn_1dn_2\langle(\vec r(n_1,t)-\vec r(n_2,t))^2\rangle$, which represents the continuous version of Eq. (\ref{gyration_radii_def}).

\section{Analytical solution in the limit $\sigma\gg1$}
\label{app: analytics}

Substituting Eq.~(\ref{eq: green_simplified}) into Eq.~(\ref{eq: mss_approx}) and introducing $\tau= \frac{\nu_0^2}{2\gamma}(t-t_0)$, we obtain:
\begin{equation}
\begin{aligned}\label{eq: D2}
\langle r^2(l)\rangle = \frac{3D}{2\pi\gamma}\int\limits_0^{\infty}dl_0\exp\left[\frac{\nu_0}{\gamma}(l-l_0)\right]\int\limits_0^{\infty}\frac{d\tau}{\tau}e^{-\tau}\left(\exp\left[-\frac{\nu_0^2(l-l_0)^2}{4\gamma^2\tau} \right]+\exp\left[-\frac{\nu_0^2(l+l_0)^2}{4\gamma^2\tau} \right]-2\exp\left[-\frac{\nu_0^2(l^2+l_0^2)}{4\gamma^2\tau} \right] \right)\\ =\frac{3D}{\pi\gamma}\int\limits_0^{\infty}dl_0\exp\left[\frac{\nu_0}{\gamma}(l-l_0)\right]\left(K_0\left[\frac{\nu_0}{\gamma}|l-l_0|\right]+K_0\left[\frac{\nu_0}{\gamma}(l+l_0)\right]-2K_0\left[\frac{\nu_0}{\gamma}\sqrt{l^2+l_0^2}\right] \right)=\frac{3D}{\pi\nu_0}f\left(\frac{l}{l_\ast}\right),
\end{aligned}
\end{equation}
where $l_\ast=\gamma/\nu_0$, $K_0$ is the modified Bessel function of the second kind, and $f(x)$ is defined as
\begin{equation}
f(x)=\int\limits_0^{\infty}dy\exp\left[x-y\right]\left(K_0\left[|x-y|\right]+K_0\left[x+y\right]-2K_0\left[\sqrt{x^2+y^2}\right] \right). \label{eq: f}
\end{equation}
%Here, we made use of the fact that, for some constant $a$,
%\begin{equation}
%    \int_{0}^{+ \infty} \frac{d \tau}{\tau} \exp\left( -\tau - \frac{a^2}{\tau} \right) = 2 K_0 \left(2 |a| \right).
%\end{equation}
%It is easy to see by factoring out $2 a$ in the argument of the exponent and then, after the substitution $ e^{x} = \tau / a$, almost immediately arriving at the standard integral representation
%\begin{equation}
%    K_0(z) = \frac{1}{2}\int_{-\infty}^{+ \infty}e^{-z \cosh{x}} dx.
%\end{equation}
After recalling the asymptotic behavior of $K_0(z) \to \sqrt{\frac{\pi}{2}}\frac{e^{-z}}{\sqrt{z}}$ for $|z| \gg 1$
%\begin{equation}
%    K_0(z) \to \sqrt{\frac{\pi}{2}}\frac{e^{-z}}{\sqrt{z}}, \label{eq: bessel_assympt}
%\end{equation}
we note that the only term in Eq. (\ref{eq: f}) that doesn't decay exponentially for $l \gg \gamma / \nu_0$ is
\begin{equation}
f(x)\approx\int\limits_0^{\infty}dy\exp\left[x-y\right]K_0\left[|x-y|\right]\approx\sqrt{2\pi x}.
\end{equation}
%where we once again used the asymptotic form of $K_0$ to arrive at the second approximate equality.
%\begin{equation}
%\langle r^2(l)\rangle\approx 3\sqrt{\frac{2}{\pi}}\frac{D\sqrt{l}}{\sqrt{\gamma\nu_0}} .\label{eq: D6}
%\end{equation}
Therefore, at $l\gg l_\ast$ we obtain expression (\ref{eq: mss_approx}) from the main text. Similary, when calculating pair correlators we arrive at
\begin{equation}
\begin{aligned}
\langle{\vec{r}( l + \Delta l, t) \cdot \vec{r}(l, t) \rangle}=\frac{3D}{\pi\gamma}\int\limits_0^{\infty}dl_0\exp\left[\frac{\nu_0}{2\gamma}(2 \cdot (l-l_0) + \Delta l)\right]
\scriptstyle \Big( K_0\left[\frac{\nu_0}{\gamma}\sqrt{(l-l_0)^2+\Delta l \cdot (l-l_0)
+ \frac{(\Delta l)^2}{2}}\right] + \\
\scriptstyle K_0\left[\frac{\nu_0}{\gamma}\sqrt{(l+l_0)^2+\Delta l \cdot (l+l_0) + \frac{(\Delta l)^2}{2}}\right]- K_0\left[\frac{\nu_0}{\gamma}\sqrt{(l^2 + l_0^2)+\Delta l \cdot (l+l_0) + \frac{(\Delta l)^2}{2}}\right]
-K_0\left[\frac{\nu_0}{\gamma}\sqrt{(l^2+l_0^2)+\Delta l \cdot (l-l_0) + \frac{(\Delta l)^2}{2}}\right]\Big).
\end{aligned}
\end{equation}
Once again, the only term that isn't exponentially suppressed for $\sigma \gg 1$ is the first one because it is the only one that features small arguments of $K_0(z)$. We note that this integral has three distinct areas of contribution: $l_0 \in (0, \ l - \omega), \ (l - \omega, \ l + \omega)\ \text{and} \ (l + \omega, \ +\infty)$, where $\omega$ controls whether $\left|(l - l_0)/l \right| \sim 1$ or not. In the former case we are allowed to perform second-order binomial expansion and use the asymptotic behavior of $K_0(z)$. Otherwise, we should check whether the contribution from $l_0 \sim l$ would be relevant

%\begin{equation}
  %  K_0\left[\frac{\nu_0}{\gamma}\sqrt{(l-l_0)^2+\Delta l \cdot (l-l_0) + \frac{(\Delta l)^2}{2}}\right] \approx K_0\left[\frac{\nu_0 \cdot |l - l_0|}{\gamma}\left(1 + \frac{\Delta l}{2(l - l_0)} + \left(\frac{\Delta l}{2(l - l_0)}\right)^2 - \frac{1}{8}\left(\frac{\Delta l}{(l - l_0)}\right)^2  \right)\right]
%\end{equation}
\begin{equation}
\begin{aligned}
    \langle{\vec{r}( l + \Delta l, t) \cdot \vec{r}(l, t) \rangle} \approx \frac{3D}{\pi \gamma} \sqrt{\frac{\pi}{2}} \int_{0}^{l - \omega}dl_0 \ \frac{\exp \left(-\frac{\nu_0}{8\gamma} \frac{(\Delta l)^2}{(l - l_0)} \right)}{\sqrt{\frac{\nu_0}{\gamma}(l - l_0)}} \cdot \left(1 +O\left(\frac{\Delta l}{l}\right) \right) + \\
    \frac{3D}{\pi \gamma}  \sqrt{\frac{\pi}{2}}\int_{l +\omega}^{\infty}dl_0 \ \frac{\exp\left( -\frac{2 \nu_0}{\gamma}(l_0 - l) + \frac{\Delta l}{\gamma}{\nu_0}\right)\cdot \exp\left( -\frac{\nu_0}{8\gamma} \frac{(\Delta l)^2}{(l_0 - l)} \right)}{\sqrt{\frac{\nu_0}{\gamma}(l_0 - l)}} \cdot \left(1 + O\left(\frac{\Delta l}{l}\right)\right) + \\
    \frac{3D}{\pi \gamma}  \int_{l - \omega}^{l + \omega}dl_0 \int_{0}^{\infty}dz \ \exp\left(\frac{\nu_0}{2\gamma}(2 \cdot (l-l_0) + \Delta l)\right) \cdot \exp \left( \cosh{z} \cdot \frac{\nu_0}{\gamma}\sqrt{(l-l_0)^2+\Delta l \cdot (l-l_0) + \frac{(\Delta l)^2}{2}}  \right).
\end{aligned}
\end{equation}
The second term has an upper bound of $3D/(2\gamma)$, which is independent of $l$. The third term is suppressed as $\omega$ grows, and also doesn't feature $l$. So, after taking the limit $\omega \to 0$ and ignoring the constant contribution, we obtain Eq. (\ref{pair_corr_approx}) from the main text.

\section{Two-sided extrusion.}\label{app:TS}

In the case of two-sided extrusion, the stochastic dynamics of loop conformation is described by Eq.~(\ref{continuous}) which should be supplemented by the zero conditions $\vec r(-N_l(t),t)= \vec r(+N_r(t),t)$ at the boundaries of the growing domain $n\in[-\nu_lt, +\nu_rt]$, where $N_l(t)=\nu_lt$ and $N_r(t)=\nu_rt$ with $\nu_l=\frac{1-q}{2}\nu_0$ and $\nu_r=\frac{1+q}{2}\nu_0$.
%representing the rates at which the LEF adds new beads at the left and right of the loop, respectively, and $q$ is the asymmetry score, introduced in Ref. [???] to characterize the  degree of asymmetry of the loop extrusion process; the choice $q=1$ corresponds to the pure  one-sided extrusion, while at  $q=0$ we deal with perfectly symmetric two-sided loop growth.
Let us pass to the new variable $l=n+\nu_Lt$. Clearly, $l\in[0,N(t)]$, where $N(t)=N_l(t)+N_r(t)=\nu_0 t$. Then Eq.~(\ref{continuous}) becomes
\begin{equation}
\label{continuum_q}
\frac{\partial \vec r(l,t)}{\partial t}=\frac{k}{\zeta} \frac{\partial^2\vec r(l,t)}{\partial l^2}-\nu_L\frac{\partial \vec r(l,t)}{\partial l}+ \frac{1}{\zeta}\vec \xi(l-\nu_Lt,t),
\end{equation}
where $\vec r(0,t)=\vec r(N(t), t)=0$. Exploiting the results reported in Ref.~\cite{Redner_2015}, we write the Green function of Eq.~(\ref{continuum_q}) as
\begin{equation}
\begin{aligned}\label{Green_function_q}
G_q(l,t;l_0,t_0)=\frac{2}{\sqrt{N(t_0)N(t)}}\exp\left[-\frac{\nu_0}{4 \gamma} \left( \frac{l^2}{N(t)} - \frac{l_0^2}{N(t_0)} \right) +\frac{q \nu_0(l-l_0)}{2\gamma} -\frac{q^2\nu_0^2(t-t_0)}{4\gamma} \right] \\ \times \sum_{j=1}^{\infty}\sin\left(\frac{j\pi l}{N(t)}\right)\sin\left(\frac{j\pi l_0}{N(t_0)}\right)
\exp\left(-\frac{j^2\pi^2\gamma (t-t_0)}{N(t_0)N(t)}\right).
\end{aligned}
\end{equation}
The gyration radius of the loop is given by expression~(\ref{gyration_app}), but now relation~(\ref{Green_function_q}) for the Green function should be used. Applying the Poisson summation formula~(\ref{Poisson}), we can  effectively evaluate $R_g^2$ numerically.

\end{widetext}

\bibliography{extruding_loops.bib}

\begin{thebibliography}{44}%
\makeatletter
\providecommand \@ifxundefined [1]{%
 \@ifx{#1\undefined}
}%
\providecommand \@ifnum [1]{%
 \ifnum #1\expandafter \@firstoftwo
 \else \expandafter \@secondoftwo
 \fi
}%
\providecommand \@ifx [1]{%
 \ifx #1\expandafter \@firstoftwo
 \else \expandafter \@secondoftwo
 \fi
}%
\providecommand \natexlab [1]{#1}%
\providecommand \enquote  [1]{``#1''}%
\providecommand \bibnamefont  [1]{#1}%
\providecommand \bibfnamefont [1]{#1}%
\providecommand \citenamefont [1]{#1}%
\providecommand \href@noop [0]{\@secondoftwo}%
\providecommand \href [0]{\begingroup \@sanitize@url \@href}%
\providecommand \@href[1]{\@@startlink{#1}\@@href}%
\providecommand \@@href[1]{\endgroup#1\@@endlink}%
\providecommand \@sanitize@url [0]{\catcode `\\12\catcode `\$12\catcode
  `\&12\catcode `\#12\catcode `\^12\catcode `\_12\catcode `\%12\relax}%
\providecommand \@@startlink[1]{}%
\providecommand \@@endlink[0]{}%
\providecommand \url  [0]{\begingroup\@sanitize@url \@url }%
\providecommand \@url [1]{\endgroup\@href {#1}{\urlprefix }}%
\providecommand \urlprefix  [0]{URL }%
\providecommand \Eprint [0]{\href }%
\providecommand \doibase [0]{https://doi.org/}%
\providecommand \selectlanguage [0]{\@gobble}%
\providecommand \bibinfo  [0]{\@secondoftwo}%
\providecommand \bibfield  [0]{\@secondoftwo}%
\providecommand \translation [1]{[#1]}%
\providecommand \BibitemOpen [0]{}%
\providecommand \bibitemStop [0]{}%
\providecommand \bibitemNoStop [0]{.\EOS\space}%
\providecommand \EOS [0]{\spacefactor3000\relax}%
\providecommand \BibitemShut  [1]{\csname bibitem#1\endcsname}%
\let\auto@bib@innerbib\@empty
%</preamble>
\bibitem [{\citenamefont {Alipour}\ and\ \citenamefont
  {Marko}(2012)}]{Alipour_2012}%
  \BibitemOpen
  \bibfield  {author} {\bibinfo {author} {\bibfnamefont {E.}~\bibnamefont
  {Alipour}}\ and\ \bibinfo {author} {\bibfnamefont {J.~F.}\ \bibnamefont
  {Marko}},\ }\bibfield  {title} {\bibinfo {title} {Self-organization of domain
  structures by dna-loop-extruding enzymes},\ }\href@noop {} {\bibfield
  {journal} {\bibinfo  {journal} {Nucleic acids research}\ }\textbf {\bibinfo
  {volume} {40}},\ \bibinfo {pages} {11202} (\bibinfo {year}
  {2012})}\BibitemShut {NoStop}%
\bibitem [{\citenamefont {Goloborodko}\ \emph {et~al.}(2016)\citenamefont
  {Goloborodko}, \citenamefont {Marko},\ and\ \citenamefont
  {Mirny}}]{Goloborodko_2016}%
  \BibitemOpen
  \bibfield  {author} {\bibinfo {author} {\bibfnamefont {A.}~\bibnamefont
  {Goloborodko}}, \bibinfo {author} {\bibfnamefont {J.~F.}\ \bibnamefont
  {Marko}},\ and\ \bibinfo {author} {\bibfnamefont {L.~A.}\ \bibnamefont
  {Mirny}},\ }\bibfield  {title} {\bibinfo {title} {Chromosome compaction by
  active loop extrusion},\ }\href@noop {} {\bibfield  {journal} {\bibinfo
  {journal} {Biophysical journal}\ }\textbf {\bibinfo {volume} {110}},\
  \bibinfo {pages} {2162} (\bibinfo {year} {2016})}\BibitemShut {NoStop}%
\bibitem [{\citenamefont {Sanborn}\ \emph {et~al.}(2015)\citenamefont
  {Sanborn}, \citenamefont {Rao}, \citenamefont {Huang}, \citenamefont
  {Durand}, \citenamefont {Huntley}, \citenamefont {Jewett}, \citenamefont
  {Bochkov}, \citenamefont {Chinnappan}, \citenamefont {Cutkosky},
  \citenamefont {Li} \emph {et~al.}}]{Sanborn_2015}%
  \BibitemOpen
  \bibfield  {author} {\bibinfo {author} {\bibfnamefont {A.~L.}\ \bibnamefont
  {Sanborn}}, \bibinfo {author} {\bibfnamefont {S.~S.}\ \bibnamefont {Rao}},
  \bibinfo {author} {\bibfnamefont {S.-C.}\ \bibnamefont {Huang}}, \bibinfo
  {author} {\bibfnamefont {N.~C.}\ \bibnamefont {Durand}}, \bibinfo {author}
  {\bibfnamefont {M.~H.}\ \bibnamefont {Huntley}}, \bibinfo {author}
  {\bibfnamefont {A.~I.}\ \bibnamefont {Jewett}}, \bibinfo {author}
  {\bibfnamefont {I.~D.}\ \bibnamefont {Bochkov}}, \bibinfo {author}
  {\bibfnamefont {D.}~\bibnamefont {Chinnappan}}, \bibinfo {author}
  {\bibfnamefont {A.}~\bibnamefont {Cutkosky}}, \bibinfo {author}
  {\bibfnamefont {J.}~\bibnamefont {Li}}, \emph {et~al.},\ }\bibfield  {title}
  {\bibinfo {title} {Chromatin extrusion explains key features of loop and
  domain formation in wild-type and engineered genomes},\ }\href@noop {}
  {\bibfield  {journal} {\bibinfo  {journal} {Proceedings of the National
  Academy of Sciences}\ }\textbf {\bibinfo {volume} {112}},\ \bibinfo {pages}
  {E6456} (\bibinfo {year} {2015})}\BibitemShut {NoStop}%
\bibitem [{\citenamefont {Nuebler}\ \emph {et~al.}(2018)\citenamefont
  {Nuebler}, \citenamefont {Fudenberg}, \citenamefont {Imakaev}, \citenamefont
  {Abdennur},\ and\ \citenamefont {Mirny}}]{Nuebler_2018}%
  \BibitemOpen
  \bibfield  {author} {\bibinfo {author} {\bibfnamefont {J.}~\bibnamefont
  {Nuebler}}, \bibinfo {author} {\bibfnamefont {G.}~\bibnamefont {Fudenberg}},
  \bibinfo {author} {\bibfnamefont {M.}~\bibnamefont {Imakaev}}, \bibinfo
  {author} {\bibfnamefont {N.}~\bibnamefont {Abdennur}},\ and\ \bibinfo
  {author} {\bibfnamefont {L.~A.}\ \bibnamefont {Mirny}},\ }\bibfield  {title}
  {\bibinfo {title} {Chromatin organization by an interplay of loop extrusion
  and compartmental segregation},\ }\href@noop {} {\bibfield  {journal}
  {\bibinfo  {journal} {Proceedings of the National Academy of Sciences}\
  }\textbf {\bibinfo {volume} {115}},\ \bibinfo {pages} {E6697} (\bibinfo
  {year} {2018})}\BibitemShut {NoStop}%
\bibitem [{\citenamefont {Gibcus}\ \emph {et~al.}(2018)\citenamefont {Gibcus},
  \citenamefont {Samejima}, \citenamefont {Goloborodko}, \citenamefont
  {Samejima}, \citenamefont {Naumova}, \citenamefont {Nuebler}, \citenamefont
  {Kanemaki}, \citenamefont {Xie}, \citenamefont {Paulson}, \citenamefont
  {Earnshaw} \emph {et~al.}}]{Gibcus_2018}%
  \BibitemOpen
  \bibfield  {author} {\bibinfo {author} {\bibfnamefont {J.~H.}\ \bibnamefont
  {Gibcus}}, \bibinfo {author} {\bibfnamefont {K.}~\bibnamefont {Samejima}},
  \bibinfo {author} {\bibfnamefont {A.}~\bibnamefont {Goloborodko}}, \bibinfo
  {author} {\bibfnamefont {I.}~\bibnamefont {Samejima}}, \bibinfo {author}
  {\bibfnamefont {N.}~\bibnamefont {Naumova}}, \bibinfo {author} {\bibfnamefont
  {J.}~\bibnamefont {Nuebler}}, \bibinfo {author} {\bibfnamefont {M.~T.}\
  \bibnamefont {Kanemaki}}, \bibinfo {author} {\bibfnamefont {L.}~\bibnamefont
  {Xie}}, \bibinfo {author} {\bibfnamefont {J.~R.}\ \bibnamefont {Paulson}},
  \bibinfo {author} {\bibfnamefont {W.~C.}\ \bibnamefont {Earnshaw}}, \emph
  {et~al.},\ }\bibfield  {title} {\bibinfo {title} {A pathway for mitotic
  chromosome formation},\ }\href@noop {} {\bibfield  {journal} {\bibinfo
  {journal} {Science}\ }\textbf {\bibinfo {volume} {359}} (\bibinfo {year}
  {2018})}\BibitemShut {NoStop}%
\bibitem [{\citenamefont {Banigan}\ \emph {et~al.}(2020)\citenamefont
  {Banigan}, \citenamefont {van~den Berg}, \citenamefont {Brand{\~a}o},
  \citenamefont {Marko},\ and\ \citenamefont {Mirny}}]{Banigan_2020b}%
  \BibitemOpen
  \bibfield  {author} {\bibinfo {author} {\bibfnamefont {E.~J.}\ \bibnamefont
  {Banigan}}, \bibinfo {author} {\bibfnamefont {A.~A.}\ \bibnamefont {van~den
  Berg}}, \bibinfo {author} {\bibfnamefont {H.~B.}\ \bibnamefont
  {Brand{\~a}o}}, \bibinfo {author} {\bibfnamefont {J.~F.}\ \bibnamefont
  {Marko}},\ and\ \bibinfo {author} {\bibfnamefont {L.~A.}\ \bibnamefont
  {Mirny}},\ }\bibfield  {title} {\bibinfo {title} {Chromosome organization by
  one-sided and two-sided loop extrusion},\ }\href@noop {} {\bibfield
  {journal} {\bibinfo  {journal} {Elife}\ }\textbf {\bibinfo {volume} {9}},\
  \bibinfo {pages} {e53558} (\bibinfo {year} {2020})}\BibitemShut {NoStop}%
\bibitem [{\citenamefont {Ganji}\ \emph {et~al.}(2018)\citenamefont {Ganji},
  \citenamefont {Shaltiel}, \citenamefont {Bisht}, \citenamefont {Kim},
  \citenamefont {Kalichava}, \citenamefont {Haering},\ and\ \citenamefont
  {Dekker}}]{Ganji_2018}%
  \BibitemOpen
  \bibfield  {author} {\bibinfo {author} {\bibfnamefont {M.}~\bibnamefont
  {Ganji}}, \bibinfo {author} {\bibfnamefont {I.~A.}\ \bibnamefont {Shaltiel}},
  \bibinfo {author} {\bibfnamefont {S.}~\bibnamefont {Bisht}}, \bibinfo
  {author} {\bibfnamefont {E.}~\bibnamefont {Kim}}, \bibinfo {author}
  {\bibfnamefont {A.}~\bibnamefont {Kalichava}}, \bibinfo {author}
  {\bibfnamefont {C.~H.}\ \bibnamefont {Haering}},\ and\ \bibinfo {author}
  {\bibfnamefont {C.}~\bibnamefont {Dekker}},\ }\bibfield  {title} {\bibinfo
  {title} {Real-time imaging of dna loop extrusion by condensin},\ }\href@noop
  {} {\bibfield  {journal} {\bibinfo  {journal} {Science}\ }\textbf {\bibinfo
  {volume} {360}},\ \bibinfo {pages} {102} (\bibinfo {year}
  {2018})}\BibitemShut {NoStop}%
\bibitem [{\citenamefont {Golfier}\ \emph {et~al.}(2020)\citenamefont
  {Golfier}, \citenamefont {Quail}, \citenamefont {Kimura},\ and\ \citenamefont
  {Brugu{\'e}s}}]{Golfier_2020}%
  \BibitemOpen
  \bibfield  {author} {\bibinfo {author} {\bibfnamefont {S.}~\bibnamefont
  {Golfier}}, \bibinfo {author} {\bibfnamefont {T.}~\bibnamefont {Quail}},
  \bibinfo {author} {\bibfnamefont {H.}~\bibnamefont {Kimura}},\ and\ \bibinfo
  {author} {\bibfnamefont {J.}~\bibnamefont {Brugu{\'e}s}},\ }\bibfield
  {title} {\bibinfo {title} {Cohesin and condensin extrude dna loops in a cell
  cycle-dependent manner},\ }\href@noop {} {\bibfield  {journal} {\bibinfo
  {journal} {Elife}\ }\textbf {\bibinfo {volume} {9}},\ \bibinfo {pages}
  {e53885} (\bibinfo {year} {2020})}\BibitemShut {NoStop}%
\bibitem [{\citenamefont {Kong}\ \emph {et~al.}(2020)\citenamefont {Kong},
  \citenamefont {Cutts}, \citenamefont {Pan}, \citenamefont {Beuron},
  \citenamefont {Kaliyappan}, \citenamefont {Xue}, \citenamefont {Morris},
  \citenamefont {Musacchio}, \citenamefont {Vannini},\ and\ \citenamefont
  {Greene}}]{Kong_2020}%
  \BibitemOpen
  \bibfield  {author} {\bibinfo {author} {\bibfnamefont {M.}~\bibnamefont
  {Kong}}, \bibinfo {author} {\bibfnamefont {E.~E.}\ \bibnamefont {Cutts}},
  \bibinfo {author} {\bibfnamefont {D.}~\bibnamefont {Pan}}, \bibinfo {author}
  {\bibfnamefont {F.}~\bibnamefont {Beuron}}, \bibinfo {author} {\bibfnamefont
  {T.}~\bibnamefont {Kaliyappan}}, \bibinfo {author} {\bibfnamefont
  {C.}~\bibnamefont {Xue}}, \bibinfo {author} {\bibfnamefont {E.~P.}\
  \bibnamefont {Morris}}, \bibinfo {author} {\bibfnamefont {A.}~\bibnamefont
  {Musacchio}}, \bibinfo {author} {\bibfnamefont {A.}~\bibnamefont {Vannini}},\
  and\ \bibinfo {author} {\bibfnamefont {E.~C.}\ \bibnamefont {Greene}},\
  }\bibfield  {title} {\bibinfo {title} {Human condensin i and ii drive
  extensive atp-dependent compaction of nucleosome-bound dna},\ }\href@noop {}
  {\bibfield  {journal} {\bibinfo  {journal} {Molecular cell}\ }\textbf
  {\bibinfo {volume} {79}},\ \bibinfo {pages} {99} (\bibinfo {year}
  {2020})}\BibitemShut {NoStop}%
\bibitem [{\citenamefont {De~Gennes}\ and\ \citenamefont
  {Gennes}(1979)}]{DeGennes_1979}%
  \BibitemOpen
  \bibfield  {author} {\bibinfo {author} {\bibfnamefont {P.-G.}\ \bibnamefont
  {De~Gennes}}\ and\ \bibinfo {author} {\bibfnamefont {P.-G.}\ \bibnamefont
  {Gennes}},\ }\href@noop {} {\emph {\bibinfo {title} {Scaling concepts in
  polymer physics}}}\ (\bibinfo  {publisher} {Cornell university press},\
  \bibinfo {year} {1979})\BibitemShut {NoStop}%
\bibitem [{\citenamefont {Doi}\ \emph {et~al.}(1988)\citenamefont {Doi},
  \citenamefont {Edwards},\ and\ \citenamefont {Edwards}}]{Doi_1988}%
  \BibitemOpen
  \bibfield  {author} {\bibinfo {author} {\bibfnamefont {M.}~\bibnamefont
  {Doi}}, \bibinfo {author} {\bibfnamefont {S.~F.}\ \bibnamefont {Edwards}},\
  and\ \bibinfo {author} {\bibfnamefont {S.~F.}\ \bibnamefont {Edwards}},\
  }\href@noop {} {\emph {\bibinfo {title} {The theory of polymer dynamics}}},\
  Vol.~\bibinfo {volume} {73}\ (\bibinfo  {publisher} {oxford university
  press},\ \bibinfo {year} {1988})\BibitemShut {NoStop}%
\bibitem [{\citenamefont {Grosberg}\ and\ \citenamefont
  {Khokhlov}(1994)}]{GKh_1994}%
  \BibitemOpen
  \bibfield  {author} {\bibinfo {author} {\bibfnamefont {A.~Y.}\ \bibnamefont
  {Grosberg}}\ and\ \bibinfo {author} {\bibfnamefont {A.}~\bibnamefont
  {Khokhlov}},\ }\href@noop {} {\emph {\bibinfo {title} {Statistical Physics of
  Macromolecules (AIP, Woodbury, NY)}}}\ (\bibinfo  {publisher} {Woodbury, NY:
  AIP Press},\ \bibinfo {year} {1994})\BibitemShut {NoStop}%
\bibitem [{\citenamefont {Grosberg}(2016)}]{Grosberg_2016}%
  \BibitemOpen
  \bibfield  {author} {\bibinfo {author} {\bibfnamefont {A.~Y.}\ \bibnamefont
  {Grosberg}},\ }\bibfield  {title} {\bibinfo {title} {Extruding loops to make
  loopy globules?},\ }\href@noop {} {\bibfield  {journal} {\bibinfo  {journal}
  {Biophysical journal}\ }\textbf {\bibinfo {volume} {110}},\ \bibinfo {pages}
  {2133} (\bibinfo {year} {2016})}\BibitemShut {NoStop}%
\bibitem [{\citenamefont {Huang}\ \emph {et~al.}(2018)\citenamefont {Huang},
  \citenamefont {Lin}, \citenamefont {Fr{\"o}mberg}, \citenamefont {Shin},
  \citenamefont {J{\"u}licher},\ and\ \citenamefont {Zaburdaev}}]{Huang_2018}%
  \BibitemOpen
  \bibfield  {author} {\bibinfo {author} {\bibfnamefont {W.}~\bibnamefont
  {Huang}}, \bibinfo {author} {\bibfnamefont {Y.~T.}\ \bibnamefont {Lin}},
  \bibinfo {author} {\bibfnamefont {D.}~\bibnamefont {Fr{\"o}mberg}}, \bibinfo
  {author} {\bibfnamefont {J.}~\bibnamefont {Shin}}, \bibinfo {author}
  {\bibfnamefont {F.}~\bibnamefont {J{\"u}licher}},\ and\ \bibinfo {author}
  {\bibfnamefont {V.}~\bibnamefont {Zaburdaev}},\ }\bibfield  {title} {\bibinfo
  {title} {Exactly solvable dynamics of forced polymer loops},\ }\href@noop {}
  {\bibfield  {journal} {\bibinfo  {journal} {New Journal of Physics}\ }\textbf
  {\bibinfo {volume} {20}},\ \bibinfo {pages} {113005} (\bibinfo {year}
  {2018})}\BibitemShut {NoStop}%
\bibitem [{\citenamefont {Vandebroek}\ and\ \citenamefont
  {Vanderzande}(2015)}]{Vandebroek_2015}%
  \BibitemOpen
  \bibfield  {author} {\bibinfo {author} {\bibfnamefont {H.}~\bibnamefont
  {Vandebroek}}\ and\ \bibinfo {author} {\bibfnamefont {C.}~\bibnamefont
  {Vanderzande}},\ }\bibfield  {title} {\bibinfo {title} {Dynamics of a polymer
  in an active and viscoelastic bath},\ }\href@noop {} {\bibfield  {journal}
  {\bibinfo  {journal} {Physical Review E}\ }\textbf {\bibinfo {volume} {92}},\
  \bibinfo {pages} {060601} (\bibinfo {year} {2015})}\BibitemShut {NoStop}%
\bibitem [{\citenamefont {Samanta}\ and\ \citenamefont
  {Chakrabarti}(2016)}]{Samanta_2016}%
  \BibitemOpen
  \bibfield  {author} {\bibinfo {author} {\bibfnamefont {N.}~\bibnamefont
  {Samanta}}\ and\ \bibinfo {author} {\bibfnamefont {R.}~\bibnamefont
  {Chakrabarti}},\ }\bibfield  {title} {\bibinfo {title} {Chain reconfiguration
  in active noise},\ }\href@noop {} {\bibfield  {journal} {\bibinfo  {journal}
  {Journal of Physics A: Mathematical and Theoretical}\ }\textbf {\bibinfo
  {volume} {49}},\ \bibinfo {pages} {195601} (\bibinfo {year}
  {2016})}\BibitemShut {NoStop}%
\bibitem [{\citenamefont {Sakaue}\ and\ \citenamefont
  {Saito}(2017)}]{Sakaue_2017}%
  \BibitemOpen
  \bibfield  {author} {\bibinfo {author} {\bibfnamefont {T.}~\bibnamefont
  {Sakaue}}\ and\ \bibinfo {author} {\bibfnamefont {T.}~\bibnamefont {Saito}},\
  }\bibfield  {title} {\bibinfo {title} {Active diffusion of model chromosomal
  loci driven by athermal noise},\ }\href@noop {} {\bibfield  {journal}
  {\bibinfo  {journal} {Soft Matter}\ }\textbf {\bibinfo {volume} {13}},\
  \bibinfo {pages} {81} (\bibinfo {year} {2017})}\BibitemShut {NoStop}%
\bibitem [{\citenamefont {Osmanovi{\'c}}\ and\ \citenamefont
  {Rabin}(2017)}]{Osmanovic_2017}%
  \BibitemOpen
  \bibfield  {author} {\bibinfo {author} {\bibfnamefont {D.}~\bibnamefont
  {Osmanovi{\'c}}}\ and\ \bibinfo {author} {\bibfnamefont {Y.}~\bibnamefont
  {Rabin}},\ }\bibfield  {title} {\bibinfo {title} {Dynamics of active rouse
  chains},\ }\href@noop {} {\bibfield  {journal} {\bibinfo  {journal} {Soft
  matter}\ }\textbf {\bibinfo {volume} {13}},\ \bibinfo {pages} {963} (\bibinfo
  {year} {2017})}\BibitemShut {NoStop}%
\bibitem [{\citenamefont {Winkler}\ \emph {et~al.}(2017)\citenamefont
  {Winkler}, \citenamefont {Elgeti},\ and\ \citenamefont
  {Gompper}}]{Winkler_2017}%
  \BibitemOpen
  \bibfield  {author} {\bibinfo {author} {\bibfnamefont {R.~G.}\ \bibnamefont
  {Winkler}}, \bibinfo {author} {\bibfnamefont {J.}~\bibnamefont {Elgeti}},\
  and\ \bibinfo {author} {\bibfnamefont {G.}~\bibnamefont {Gompper}},\
  }\bibfield  {title} {\bibinfo {title} {Active polymers emergent
  conformational and dynamical properties: a brief review},\ }\href@noop {}
  {\bibfield  {journal} {\bibinfo  {journal} {Journal of the Physical Society
  of Japan}\ }\textbf {\bibinfo {volume} {86}},\ \bibinfo {pages} {101014}
  (\bibinfo {year} {2017})}\BibitemShut {NoStop}%
\bibitem [{\citenamefont {L{\"o}wen}(2018)}]{Lowen_2018}%
  \BibitemOpen
  \bibfield  {author} {\bibinfo {author} {\bibfnamefont {H.}~\bibnamefont
  {L{\"o}wen}},\ }\bibfield  {title} {\bibinfo {title} {Active colloidal
  molecules},\ }\href@noop {} {\bibfield  {journal} {\bibinfo  {journal} {EPL
  (Europhysics Letters)}\ }\textbf {\bibinfo {volume} {121}},\ \bibinfo {pages}
  {58001} (\bibinfo {year} {2018})}\BibitemShut {NoStop}%
\bibitem [{\citenamefont {Osmanovi{\'c}}(2018)}]{Osmanovic_2018}%
  \BibitemOpen
  \bibfield  {author} {\bibinfo {author} {\bibfnamefont {D.}~\bibnamefont
  {Osmanovi{\'c}}},\ }\bibfield  {title} {\bibinfo {title} {Properties of rouse
  polymers with actively driven regions},\ }\href@noop {} {\bibfield  {journal}
  {\bibinfo  {journal} {The Journal of chemical physics}\ }\textbf {\bibinfo
  {volume} {149}},\ \bibinfo {pages} {164911} (\bibinfo {year}
  {2018})}\BibitemShut {NoStop}%
\bibitem [{\citenamefont {Chaki}\ and\ \citenamefont
  {Chakrabarti}(2019)}]{Chaki_2019}%
  \BibitemOpen
  \bibfield  {author} {\bibinfo {author} {\bibfnamefont {S.}~\bibnamefont
  {Chaki}}\ and\ \bibinfo {author} {\bibfnamefont {R.}~\bibnamefont
  {Chakrabarti}},\ }\bibfield  {title} {\bibinfo {title} {Enhanced diffusion,
  swelling, and slow reconfiguration of a single chain in non-gaussian active
  bath},\ }\href@noop {} {\bibfield  {journal} {\bibinfo  {journal} {The
  Journal of chemical physics}\ }\textbf {\bibinfo {volume} {150}},\ \bibinfo
  {pages} {094902} (\bibinfo {year} {2019})}\BibitemShut {NoStop}%
\bibitem [{\citenamefont {Put}\ \emph {et~al.}(2019)\citenamefont {Put},
  \citenamefont {Sakaue},\ and\ \citenamefont {Vanderzande}}]{Put_2019}%
  \BibitemOpen
  \bibfield  {author} {\bibinfo {author} {\bibfnamefont {S.}~\bibnamefont
  {Put}}, \bibinfo {author} {\bibfnamefont {T.}~\bibnamefont {Sakaue}},\ and\
  \bibinfo {author} {\bibfnamefont {C.}~\bibnamefont {Vanderzande}},\
  }\bibfield  {title} {\bibinfo {title} {Active dynamics and spatially coherent
  motion in chromosomes subject to enzymatic force dipoles},\ }\href@noop {}
  {\bibfield  {journal} {\bibinfo  {journal} {Physical Review E}\ }\textbf
  {\bibinfo {volume} {99}},\ \bibinfo {pages} {032421} (\bibinfo {year}
  {2019})}\BibitemShut {NoStop}%
\bibitem [{\citenamefont {Anand}\ and\ \citenamefont
  {Singh}(2020)}]{Anand_2020}%
  \BibitemOpen
  \bibfield  {author} {\bibinfo {author} {\bibfnamefont {S.~K.}\ \bibnamefont
  {Anand}}\ and\ \bibinfo {author} {\bibfnamefont {S.~P.}\ \bibnamefont
  {Singh}},\ }\bibfield  {title} {\bibinfo {title} {Conformation and dynamics
  of a self-avoiding active flexible polymer},\ }\href@noop {} {\bibfield
  {journal} {\bibinfo  {journal} {Physical Review E}\ }\textbf {\bibinfo
  {volume} {101}},\ \bibinfo {pages} {030501} (\bibinfo {year}
  {2020})}\BibitemShut {NoStop}%
\bibitem [{\citenamefont {Dekker}\ \emph {et~al.}(2002)\citenamefont {Dekker},
  \citenamefont {Rippe}, \citenamefont {Dekker},\ and\ \citenamefont
  {Kleckner}}]{Dekker_2002}%
  \BibitemOpen
  \bibfield  {author} {\bibinfo {author} {\bibfnamefont {J.}~\bibnamefont
  {Dekker}}, \bibinfo {author} {\bibfnamefont {K.}~\bibnamefont {Rippe}},
  \bibinfo {author} {\bibfnamefont {M.}~\bibnamefont {Dekker}},\ and\ \bibinfo
  {author} {\bibfnamefont {N.}~\bibnamefont {Kleckner}},\ }\bibfield  {title}
  {\bibinfo {title} {Capturing chromosome conformation},\ }\href@noop {}
  {\bibfield  {journal} {\bibinfo  {journal} {science}\ }\textbf {\bibinfo
  {volume} {295}},\ \bibinfo {pages} {1306} (\bibinfo {year}
  {2002})}\BibitemShut {NoStop}%
\bibitem [{\citenamefont {Bolzer}\ \emph {et~al.}(2005)\citenamefont {Bolzer},
  \citenamefont {Kreth}, \citenamefont {Solovei}, \citenamefont {Koehler},
  \citenamefont {Saracoglu}, \citenamefont {Fauth}, \citenamefont {M{\"u}ller},
  \citenamefont {Eils}, \citenamefont {Cremer}, \citenamefont {Speicher} \emph
  {et~al.}}]{Bolzer_2005}%
  \BibitemOpen
  \bibfield  {author} {\bibinfo {author} {\bibfnamefont {A.}~\bibnamefont
  {Bolzer}}, \bibinfo {author} {\bibfnamefont {G.}~\bibnamefont {Kreth}},
  \bibinfo {author} {\bibfnamefont {I.}~\bibnamefont {Solovei}}, \bibinfo
  {author} {\bibfnamefont {D.}~\bibnamefont {Koehler}}, \bibinfo {author}
  {\bibfnamefont {K.}~\bibnamefont {Saracoglu}}, \bibinfo {author}
  {\bibfnamefont {C.}~\bibnamefont {Fauth}}, \bibinfo {author} {\bibfnamefont
  {S.}~\bibnamefont {M{\"u}ller}}, \bibinfo {author} {\bibfnamefont
  {R.}~\bibnamefont {Eils}}, \bibinfo {author} {\bibfnamefont {C.}~\bibnamefont
  {Cremer}}, \bibinfo {author} {\bibfnamefont {M.~R.}\ \bibnamefont
  {Speicher}}, \emph {et~al.},\ }\bibfield  {title} {\bibinfo {title}
  {Three-dimensional maps of all chromosomes in human male fibroblast nuclei
  and prometaphase rosettes},\ }\href@noop {} {\bibfield  {journal} {\bibinfo
  {journal} {PLoS Biol}\ }\textbf {\bibinfo {volume} {3}},\ \bibinfo {pages}
  {e157} (\bibinfo {year} {2005})}\BibitemShut {NoStop}%
\bibitem [{\citenamefont {Lieberman-Aiden}\ \emph {et~al.}(2009)\citenamefont
  {Lieberman-Aiden}, \citenamefont {Van~Berkum}, \citenamefont {Williams},
  \citenamefont {Imakaev}, \citenamefont {Ragoczy}, \citenamefont {Telling},
  \citenamefont {Amit}, \citenamefont {Lajoie}, \citenamefont {Sabo},
  \citenamefont {Dorschner} \emph {et~al.}}]{Lieberman_Aiden_2009}%
  \BibitemOpen
  \bibfield  {author} {\bibinfo {author} {\bibfnamefont {E.}~\bibnamefont
  {Lieberman-Aiden}}, \bibinfo {author} {\bibfnamefont {N.~L.}\ \bibnamefont
  {Van~Berkum}}, \bibinfo {author} {\bibfnamefont {L.}~\bibnamefont
  {Williams}}, \bibinfo {author} {\bibfnamefont {M.}~\bibnamefont {Imakaev}},
  \bibinfo {author} {\bibfnamefont {T.}~\bibnamefont {Ragoczy}}, \bibinfo
  {author} {\bibfnamefont {A.}~\bibnamefont {Telling}}, \bibinfo {author}
  {\bibfnamefont {I.}~\bibnamefont {Amit}}, \bibinfo {author} {\bibfnamefont
  {B.~R.}\ \bibnamefont {Lajoie}}, \bibinfo {author} {\bibfnamefont {P.~J.}\
  \bibnamefont {Sabo}}, \bibinfo {author} {\bibfnamefont {M.~O.}\ \bibnamefont
  {Dorschner}}, \emph {et~al.},\ }\bibfield  {title} {\bibinfo {title}
  {Comprehensive mapping of long-range interactions reveals folding principles
  of the human genome},\ }\href@noop {} {\bibfield  {journal} {\bibinfo
  {journal} {science}\ }\textbf {\bibinfo {volume} {326}},\ \bibinfo {pages}
  {289} (\bibinfo {year} {2009})}\BibitemShut {NoStop}%
\bibitem [{\citenamefont {Joyce}\ \emph {et~al.}(2012)\citenamefont {Joyce},
  \citenamefont {Williams}, \citenamefont {Xie} \emph {et~al.}}]{Joyce_2012}%
  \BibitemOpen
  \bibfield  {author} {\bibinfo {author} {\bibfnamefont {E.~F.}\ \bibnamefont
  {Joyce}}, \bibinfo {author} {\bibfnamefont {B.~R.}\ \bibnamefont {Williams}},
  \bibinfo {author} {\bibfnamefont {T.}~\bibnamefont {Xie}}, \emph {et~al.},\
  }\bibfield  {title} {\bibinfo {title} {Identification of genes that promote
  or antagonize somatic homolog pairing using a high-throughput fish--based
  screen},\ }\href@noop {} {\bibfield  {journal} {\bibinfo  {journal} {PLoS
  Genet}\ }\textbf {\bibinfo {volume} {8}},\ \bibinfo {pages} {e1002667}
  (\bibinfo {year} {2012})}\BibitemShut {NoStop}%
\bibitem [{\citenamefont {Nagano}\ \emph {et~al.}(2013)\citenamefont {Nagano},
  \citenamefont {Lubling}, \citenamefont {Stevens}, \citenamefont
  {Schoenfelder}, \citenamefont {Yaffe}, \citenamefont {Dean}, \citenamefont
  {Laue}, \citenamefont {Tanay},\ and\ \citenamefont {Fraser}}]{Nagano_2013}%
  \BibitemOpen
  \bibfield  {author} {\bibinfo {author} {\bibfnamefont {T.}~\bibnamefont
  {Nagano}}, \bibinfo {author} {\bibfnamefont {Y.}~\bibnamefont {Lubling}},
  \bibinfo {author} {\bibfnamefont {T.~J.}\ \bibnamefont {Stevens}}, \bibinfo
  {author} {\bibfnamefont {S.}~\bibnamefont {Schoenfelder}}, \bibinfo {author}
  {\bibfnamefont {E.}~\bibnamefont {Yaffe}}, \bibinfo {author} {\bibfnamefont
  {W.}~\bibnamefont {Dean}}, \bibinfo {author} {\bibfnamefont {E.~D.}\
  \bibnamefont {Laue}}, \bibinfo {author} {\bibfnamefont {A.}~\bibnamefont
  {Tanay}},\ and\ \bibinfo {author} {\bibfnamefont {P.}~\bibnamefont
  {Fraser}},\ }\bibfield  {title} {\bibinfo {title} {Single-cell hi-c reveals
  cell-to-cell variability in chromosome structure},\ }\href@noop {} {\bibfield
   {journal} {\bibinfo  {journal} {Nature}\ }\textbf {\bibinfo {volume}
  {502}},\ \bibinfo {pages} {59} (\bibinfo {year} {2013})}\BibitemShut
  {NoStop}%
\bibitem [{\citenamefont {Shachar}\ \emph {et~al.}(2015)\citenamefont
  {Shachar}, \citenamefont {Voss}, \citenamefont {Pegoraro}, \citenamefont
  {Sciascia},\ and\ \citenamefont {Misteli}}]{Shachar_2015}%
  \BibitemOpen
  \bibfield  {author} {\bibinfo {author} {\bibfnamefont {S.}~\bibnamefont
  {Shachar}}, \bibinfo {author} {\bibfnamefont {T.~C.}\ \bibnamefont {Voss}},
  \bibinfo {author} {\bibfnamefont {G.}~\bibnamefont {Pegoraro}}, \bibinfo
  {author} {\bibfnamefont {N.}~\bibnamefont {Sciascia}},\ and\ \bibinfo
  {author} {\bibfnamefont {T.}~\bibnamefont {Misteli}},\ }\bibfield  {title}
  {\bibinfo {title} {Identification of gene positioning factors using
  high-throughput imaging mapping},\ }\href@noop {} {\bibfield  {journal}
  {\bibinfo  {journal} {Cell}\ }\textbf {\bibinfo {volume} {162}},\ \bibinfo
  {pages} {911} (\bibinfo {year} {2015})}\BibitemShut {NoStop}%
\bibitem [{\citenamefont {Fraser}\ \emph {et~al.}(2015)\citenamefont {Fraser},
  \citenamefont {Williamson}, \citenamefont {Bickmore},\ and\ \citenamefont
  {Dostie}}]{Fraser_2015}%
  \BibitemOpen
  \bibfield  {author} {\bibinfo {author} {\bibfnamefont {J.}~\bibnamefont
  {Fraser}}, \bibinfo {author} {\bibfnamefont {I.}~\bibnamefont {Williamson}},
  \bibinfo {author} {\bibfnamefont {W.~A.}\ \bibnamefont {Bickmore}},\ and\
  \bibinfo {author} {\bibfnamefont {J.}~\bibnamefont {Dostie}},\ }\bibfield
  {title} {\bibinfo {title} {An overview of genome organization and how we got
  there: from fish to hi-c},\ }\href@noop {} {\bibfield  {journal} {\bibinfo
  {journal} {Microbiology and Molecular Biology Reviews}\ }\textbf {\bibinfo
  {volume} {79}},\ \bibinfo {pages} {347} (\bibinfo {year} {2015})}\BibitemShut
  {NoStop}%
\bibitem [{\citenamefont {Kind}\ \emph {et~al.}(2015)\citenamefont {Kind},
  \citenamefont {Pagie}, \citenamefont {de~Vries}, \citenamefont {Nahidiazar},
  \citenamefont {Dey}, \citenamefont {Bienko}, \citenamefont {Zhan},
  \citenamefont {Lajoie}, \citenamefont {de~Graaf}, \citenamefont {Amendola}
  \emph {et~al.}}]{Kind_2015}%
  \BibitemOpen
  \bibfield  {author} {\bibinfo {author} {\bibfnamefont {J.}~\bibnamefont
  {Kind}}, \bibinfo {author} {\bibfnamefont {L.}~\bibnamefont {Pagie}},
  \bibinfo {author} {\bibfnamefont {S.~S.}\ \bibnamefont {de~Vries}}, \bibinfo
  {author} {\bibfnamefont {L.}~\bibnamefont {Nahidiazar}}, \bibinfo {author}
  {\bibfnamefont {S.~S.}\ \bibnamefont {Dey}}, \bibinfo {author} {\bibfnamefont
  {M.}~\bibnamefont {Bienko}}, \bibinfo {author} {\bibfnamefont
  {Y.}~\bibnamefont {Zhan}}, \bibinfo {author} {\bibfnamefont {B.}~\bibnamefont
  {Lajoie}}, \bibinfo {author} {\bibfnamefont {C.~A.}\ \bibnamefont
  {de~Graaf}}, \bibinfo {author} {\bibfnamefont {M.}~\bibnamefont {Amendola}},
  \emph {et~al.},\ }\bibfield  {title} {\bibinfo {title} {Genome-wide maps of
  nuclear lamina interactions in single human cells},\ }\href@noop {}
  {\bibfield  {journal} {\bibinfo  {journal} {Cell}\ }\textbf {\bibinfo
  {volume} {163}},\ \bibinfo {pages} {134} (\bibinfo {year}
  {2015})}\BibitemShut {NoStop}%
\bibitem [{\citenamefont {Van~Berkum}\ \emph {et~al.}(2010)\citenamefont
  {Van~Berkum}, \citenamefont {Lieberman-Aiden}, \citenamefont {Williams},
  \citenamefont {Imakaev}, \citenamefont {Gnirke}, \citenamefont {Mirny},
  \citenamefont {Dekker},\ and\ \citenamefont {Lander}}]{Berkum_2010}%
  \BibitemOpen
  \bibfield  {author} {\bibinfo {author} {\bibfnamefont {N.~L.}\ \bibnamefont
  {Van~Berkum}}, \bibinfo {author} {\bibfnamefont {E.}~\bibnamefont
  {Lieberman-Aiden}}, \bibinfo {author} {\bibfnamefont {L.}~\bibnamefont
  {Williams}}, \bibinfo {author} {\bibfnamefont {M.}~\bibnamefont {Imakaev}},
  \bibinfo {author} {\bibfnamefont {A.}~\bibnamefont {Gnirke}}, \bibinfo
  {author} {\bibfnamefont {L.~A.}\ \bibnamefont {Mirny}}, \bibinfo {author}
  {\bibfnamefont {J.}~\bibnamefont {Dekker}},\ and\ \bibinfo {author}
  {\bibfnamefont {E.~S.}\ \bibnamefont {Lander}},\ }\bibfield  {title}
  {\bibinfo {title} {Hi-c: a method to study the three-dimensional architecture
  of genomes.},\ }\href@noop {} {\bibfield  {journal} {\bibinfo  {journal}
  {JoVE (Journal of Visualized Experiments)}\ ,\ \bibinfo {pages} {e1869}}
  (\bibinfo {year} {2010})}\BibitemShut {NoStop}%
\bibitem [{\citenamefont {Oudelaar}\ \emph {et~al.}(2018)\citenamefont
  {Oudelaar}, \citenamefont {Davies}, \citenamefont {Hanssen}, \citenamefont
  {Telenius}, \citenamefont {Schwessinger}, \citenamefont {Liu}, \citenamefont
  {Brown}, \citenamefont {Downes}, \citenamefont {Chiariello}, \citenamefont
  {Bianco} \emph {et~al.}}]{Oudelaar_2018}%
  \BibitemOpen
  \bibfield  {author} {\bibinfo {author} {\bibfnamefont {A.~M.}\ \bibnamefont
  {Oudelaar}}, \bibinfo {author} {\bibfnamefont {J.~O.}\ \bibnamefont
  {Davies}}, \bibinfo {author} {\bibfnamefont {L.~L.}\ \bibnamefont {Hanssen}},
  \bibinfo {author} {\bibfnamefont {J.~M.}\ \bibnamefont {Telenius}}, \bibinfo
  {author} {\bibfnamefont {R.}~\bibnamefont {Schwessinger}}, \bibinfo {author}
  {\bibfnamefont {Y.}~\bibnamefont {Liu}}, \bibinfo {author} {\bibfnamefont
  {J.~M.}\ \bibnamefont {Brown}}, \bibinfo {author} {\bibfnamefont {D.~J.}\
  \bibnamefont {Downes}}, \bibinfo {author} {\bibfnamefont {A.~M.}\
  \bibnamefont {Chiariello}}, \bibinfo {author} {\bibfnamefont
  {S.}~\bibnamefont {Bianco}}, \emph {et~al.},\ }\bibfield  {title} {\bibinfo
  {title} {Single-allele chromatin interactions identify regulatory hubs in
  dynamic compartmentalized domains},\ }\href@noop {} {\bibfield  {journal}
  {\bibinfo  {journal} {Nature genetics}\ }\textbf {\bibinfo {volume} {50}},\
  \bibinfo {pages} {1744} (\bibinfo {year} {2018})}\BibitemShut {NoStop}%
\bibitem [{\citenamefont {Tavares-Cadete}\ \emph {et~al.}(2020)\citenamefont
  {Tavares-Cadete}, \citenamefont {Norouzi}, \citenamefont {Dekker},
  \citenamefont {Liu},\ and\ \citenamefont {Dekker}}]{Tavares_Cadete_2020}%
  \BibitemOpen
  \bibfield  {author} {\bibinfo {author} {\bibfnamefont {F.}~\bibnamefont
  {Tavares-Cadete}}, \bibinfo {author} {\bibfnamefont {D.}~\bibnamefont
  {Norouzi}}, \bibinfo {author} {\bibfnamefont {B.}~\bibnamefont {Dekker}},
  \bibinfo {author} {\bibfnamefont {Y.}~\bibnamefont {Liu}},\ and\ \bibinfo
  {author} {\bibfnamefont {J.}~\bibnamefont {Dekker}},\ }\bibfield  {title}
  {\bibinfo {title} {Multi-contact 3c reveals that the human genome during
  interphase is largely not entangled},\ }\href@noop {} {\bibfield  {journal}
  {\bibinfo  {journal} {Nature Structural \& Molecular Biology}\ }\textbf
  {\bibinfo {volume} {27}},\ \bibinfo {pages} {1105} (\bibinfo {year}
  {2020})}\BibitemShut {NoStop}%
\bibitem [{\citenamefont {Oomen}\ \emph {et~al.}(2020)\citenamefont {Oomen},
  \citenamefont {Hedger}, \citenamefont {Watts},\ and\ \citenamefont
  {Dekker}}]{Oomen_2020}%
  \BibitemOpen
  \bibfield  {author} {\bibinfo {author} {\bibfnamefont {M.~E.}\ \bibnamefont
  {Oomen}}, \bibinfo {author} {\bibfnamefont {A.~K.}\ \bibnamefont {Hedger}},
  \bibinfo {author} {\bibfnamefont {J.~K.}\ \bibnamefont {Watts}},\ and\
  \bibinfo {author} {\bibfnamefont {J.}~\bibnamefont {Dekker}},\ }\bibfield
  {title} {\bibinfo {title} {Detecting chromatin interactions between and along
  sister chromatids with sisterc},\ }\href@noop {} {\bibfield  {journal}
  {\bibinfo  {journal} {Nature Methods}\ }\textbf {\bibinfo {volume} {17}},\
  \bibinfo {pages} {1002} (\bibinfo {year} {2020})}\BibitemShut {NoStop}%
\bibitem [{\citenamefont {Krietenstein}\ \emph {et~al.}(2020)\citenamefont
  {Krietenstein}, \citenamefont {Abraham}, \citenamefont {Venev}, \citenamefont
  {Abdennur}, \citenamefont {Gibcus}, \citenamefont {Hsieh}, \citenamefont
  {Parsi}, \citenamefont {Yang}, \citenamefont {Maehr}, \citenamefont {Mirny}
  \emph {et~al.}}]{Krietenstein_2020}%
  \BibitemOpen
  \bibfield  {author} {\bibinfo {author} {\bibfnamefont {N.}~\bibnamefont
  {Krietenstein}}, \bibinfo {author} {\bibfnamefont {S.}~\bibnamefont
  {Abraham}}, \bibinfo {author} {\bibfnamefont {S.~V.}\ \bibnamefont {Venev}},
  \bibinfo {author} {\bibfnamefont {N.}~\bibnamefont {Abdennur}}, \bibinfo
  {author} {\bibfnamefont {J.}~\bibnamefont {Gibcus}}, \bibinfo {author}
  {\bibfnamefont {T.-H.~S.}\ \bibnamefont {Hsieh}}, \bibinfo {author}
  {\bibfnamefont {K.~M.}\ \bibnamefont {Parsi}}, \bibinfo {author}
  {\bibfnamefont {L.}~\bibnamefont {Yang}}, \bibinfo {author} {\bibfnamefont
  {R.}~\bibnamefont {Maehr}}, \bibinfo {author} {\bibfnamefont {L.~A.}\
  \bibnamefont {Mirny}}, \emph {et~al.},\ }\bibfield  {title} {\bibinfo {title}
  {Ultrastructural details of mammalian chromosome architecture},\ }\href@noop
  {} {\bibfield  {journal} {\bibinfo  {journal} {Molecular cell}\ }\textbf
  {\bibinfo {volume} {78}},\ \bibinfo {pages} {554} (\bibinfo {year}
  {2020})}\BibitemShut {NoStop}%
\bibitem [{\citenamefont {Riggs}(1990)}]{Riggs_1990}%
  \BibitemOpen
  \bibfield  {author} {\bibinfo {author} {\bibfnamefont {A.}~\bibnamefont
  {Riggs}},\ }\bibfield  {title} {\bibinfo {title} {Dna methylation and late
  replication probably aid cell memory, and type i dna reeling could aid
  chromosome folding and enhancer function},\ }\href@noop {} {\bibfield
  {journal} {\bibinfo  {journal} {Philosophical Transactions of the Royal
  Society of London. B, Biological Sciences}\ }\textbf {\bibinfo {volume}
  {326}},\ \bibinfo {pages} {285} (\bibinfo {year} {1990})}\BibitemShut
  {NoStop}%
\bibitem [{\citenamefont {Risken}(1996)}]{Risken_1996}%
  \BibitemOpen
  \bibfield  {author} {\bibinfo {author} {\bibfnamefont {H.}~\bibnamefont
  {Risken}},\ }\bibfield  {title} {\bibinfo {title} {Fokker-planck equation},\
  }in\ \href@noop {} {\emph {\bibinfo {booktitle} {The Fokker-Planck
  Equation}}}\ (\bibinfo  {publisher} {Springer},\ \bibinfo {year} {1996})\
  pp.\ \bibinfo {pages} {63--95}\BibitemShut {NoStop}%
\bibitem [{\citenamefont {Chupeau}\ \emph {et~al.}(2015)\citenamefont
  {Chupeau}, \citenamefont {B{\'e}nichou},\ and\ \citenamefont
  {Redner}}]{Redner_2015}%
  \BibitemOpen
  \bibfield  {author} {\bibinfo {author} {\bibfnamefont {M.}~\bibnamefont
  {Chupeau}}, \bibinfo {author} {\bibfnamefont {O.}~\bibnamefont
  {B{\'e}nichou}},\ and\ \bibinfo {author} {\bibfnamefont {S.}~\bibnamefont
  {Redner}},\ }\bibfield  {title} {\bibinfo {title} {Optimal strategy to
  capture a skittish lamb wandering near a precipice},\ }\href@noop {}
  {\bibfield  {journal} {\bibinfo  {journal} {Journal of Statistical Mechanics:
  Theory and Experiment}\ }\textbf {\bibinfo {volume} {2015}},\ \bibinfo
  {pages} {P06026} (\bibinfo {year} {2015})}\BibitemShut {NoStop}%
\bibitem [{\citenamefont {Davidson}\ \emph {et~al.}(2019)\citenamefont
  {Davidson}, \citenamefont {Bauer}, \citenamefont {Goetz}, \citenamefont
  {Tang}, \citenamefont {Wutz},\ and\ \citenamefont {Peters}}]{Davidson_2019}%
  \BibitemOpen
  \bibfield  {author} {\bibinfo {author} {\bibfnamefont {I.~F.}\ \bibnamefont
  {Davidson}}, \bibinfo {author} {\bibfnamefont {B.}~\bibnamefont {Bauer}},
  \bibinfo {author} {\bibfnamefont {D.}~\bibnamefont {Goetz}}, \bibinfo
  {author} {\bibfnamefont {W.}~\bibnamefont {Tang}}, \bibinfo {author}
  {\bibfnamefont {G.}~\bibnamefont {Wutz}},\ and\ \bibinfo {author}
  {\bibfnamefont {J.-M.}\ \bibnamefont {Peters}},\ }\bibfield  {title}
  {\bibinfo {title} {Dna loop extrusion by human cohesin},\ }\href@noop {}
  {\bibfield  {journal} {\bibinfo  {journal} {Science}\ }\textbf {\bibinfo
  {volume} {366}},\ \bibinfo {pages} {1338} (\bibinfo {year}
  {2019})}\BibitemShut {NoStop}%
\bibitem [{\citenamefont {Kim}\ \emph {et~al.}(2019)\citenamefont {Kim},
  \citenamefont {Shi}, \citenamefont {Zhang}, \citenamefont {Finkelstein},\
  and\ \citenamefont {Yu}}]{Kim_2019}%
  \BibitemOpen
  \bibfield  {author} {\bibinfo {author} {\bibfnamefont {Y.}~\bibnamefont
  {Kim}}, \bibinfo {author} {\bibfnamefont {Z.}~\bibnamefont {Shi}}, \bibinfo
  {author} {\bibfnamefont {H.}~\bibnamefont {Zhang}}, \bibinfo {author}
  {\bibfnamefont {I.~J.}\ \bibnamefont {Finkelstein}},\ and\ \bibinfo {author}
  {\bibfnamefont {H.}~\bibnamefont {Yu}},\ }\bibfield  {title} {\bibinfo
  {title} {Human cohesin compacts dna by loop extrusion},\ }\href@noop {}
  {\bibfield  {journal} {\bibinfo  {journal} {Science}\ }\textbf {\bibinfo
  {volume} {366}},\ \bibinfo {pages} {1345} (\bibinfo {year}
  {2019})}\BibitemShut {NoStop}%
\bibitem [{\citenamefont {Banigan}\ and\ \citenamefont
  {Mirny}(2019)}]{Banigan_2019}%
  \BibitemOpen
  \bibfield  {author} {\bibinfo {author} {\bibfnamefont {E.~J.}\ \bibnamefont
  {Banigan}}\ and\ \bibinfo {author} {\bibfnamefont {L.~A.}\ \bibnamefont
  {Mirny}},\ }\bibfield  {title} {\bibinfo {title} {Limits of chromosome
  compaction by loop-extruding motors},\ }\href@noop {} {\bibfield  {journal}
  {\bibinfo  {journal} {Physical Review X}\ }\textbf {\bibinfo {volume} {9}},\
  \bibinfo {pages} {031007} (\bibinfo {year} {2019})}\BibitemShut {NoStop}%
\bibitem [{\citenamefont {Banigan}\ and\ \citenamefont
  {Mirny}(2020)}]{Banigan_2020a}%
  \BibitemOpen
  \bibfield  {author} {\bibinfo {author} {\bibfnamefont {E.~J.}\ \bibnamefont
  {Banigan}}\ and\ \bibinfo {author} {\bibfnamefont {L.~A.}\ \bibnamefont
  {Mirny}},\ }\bibfield  {title} {\bibinfo {title} {Loop extrusion: theory
  meets single-molecule experiments},\ }\href@noop {} {\bibfield  {journal}
  {\bibinfo  {journal} {Current opinion in cell biology}\ }\textbf {\bibinfo
  {volume} {64}},\ \bibinfo {pages} {124} (\bibinfo {year} {2020})}\BibitemShut
  {NoStop}%
\end{thebibliography}%

\end{document}